\documentclass[default,iicol]{sn-jnl}

\usepackage{url}            
\usepackage{booktabs}       
\usepackage{tabularx}
\usepackage{multirow}
\usepackage{threeparttable}
\usepackage{subfigure}
\RequirePackage[super]{nth}
\newcommand{\angstrom}{\textup{\AA}}

\jyear{2022}%

\theoremstyle{thmstyleone}%
\theoremstyle{thmstyletwo}%

\theoremstyle{thmstylethree}%
\newcommand{\model}{ViSNet}

\begin{document}

\title[Article Title]{ViSNet: an equivariant geometry-enhanced graph neural network with vector-scalar interactive message passing for molecules}








\author[1,2,3]{\fnm{Yusong} \sur{Wang}}
\equalcont{These authors contributed equally to this work.}

\author[2,3]{\fnm{Shaoning} \sur{Li}}
\equalcont{These authors contributed equally to this work.}

\author[4,5,2,3]{\fnm{Xinheng} \sur{He}}

\author[6,2,3]{\fnm{Mingyu} \sur{Li}}

\author[2]{\fnm{Zun} \sur{Wang}}

\author[1]{\fnm{Nanning} \sur{Zheng}}

\author*[2]{\fnm{Bin} \sur{Shao}}\email{binshao@microsoft.com (B. S.)}

\author[2]{\fnm{Tie-Yan} \sur{Liu}}

\author*[2]{\fnm{Tong} \sur{Wang}}\email{\\watong@microsoft.com (T. W., Lead Contact)}

\affil[1]{\orgdiv{Institute of Artificial Intelligence and Robotics}, \orgname{Xi'an Jiaotong University}, \orgaddress{\city{Xi'an}, \postcode{710049}, \country{China}}}

\affil[2]{\orgname{Microsoft Research AI4Science}, \orgaddress{\city{Beijing}, \postcode{100080}, \country{China}}}

\affil[3]{\orgname{Work done during an internship at Microsoft Research AI4Science}, \orgaddress{\city{Beijing}, \postcode{100080}, \country{China}}}

\affil[4]{\orgname{The CAS Key Laboratory of Receptor Research and State Key Laboratory of Drug Research, Shanghai Institute of Materia Medica, Chinese Academy of Sciences}, \city{Shanghai}, \postcode{201203},  \country{China}}

\affil[5]{\orgname{University of Chinese Academy of Sciences}, \city{Beijing}, \postcode{100049},  \country{China}}

\affil[6]{\orgname{Medicinal Chemistry and Bioinformatics Center, Schoold of Medicine, Shanghai Jiaotong University}, \city{Shanghai}, \postcode{200025},  \country{China}}


\abstract{
Geometric deep learning has been revolutionizing the molecular modeling field.
Despite the state-of-the-art neural network models are approaching \emph{ab initio} accuracy for molecular property prediction, their applications, such as drug discovery and molecular dynamics (MD) simulation, have been hindered by insufficient utilization of geometric information and high computational costs.  
Here we propose an equivariant geometry-enhanced graph neural network called ViSNet, which elegantly extracts geometric features and efficiently models molecular structures with low computational costs.
Our proposed \model~outperforms state-of-the-art approaches on multiple MD benchmarks, including MD17, revised MD17 and MD22, and achieves excellent chemical property prediction on QM9 and Molecule3D datasets. Additionally, \model~achieved the top winners of PCQM4Mv2 track in the OGB-LCS@NeurIPS2022 competition.
Furthermore, through a series of simulations and case studies, \model~can efficiently explore the conformational space and provide reasonable interpretability to map geometric representations to molecular structures.  
}

\keywords{Geometric Deep Learning Potential; Equivariant Graph Neural Network; Molecular Modeling}

\maketitle

\section{Introduction}\label{sec:intro}
Molecular modeling plays a crucial role in modern scientific and engineering fields, aiding in the understanding of chemical reactions, facilitating new drug development, and driving scientific and technological advancements~\cite{chow20129,singh2020molecular,lu2021activation,li2021exploring}.
One commonly used method in molecular modeling is density functional theory (DFT). 
DFT enables accurate calculations of energy, forces, and other chemical properties of molecules~\cite{kohn1965self,marx2009ab}.
However, due to the large computational requirements, DFT calculations often demand significant computational resources and time, particularly for large molecular systems or high-precision calculations. 
Machine learning (ML) offers an alternative solution by learning from reference data with \emph{ab initio} accuracy and high computational efficiency \cite{christensen2020fchl, bartok2010gaussian}. 
Behler and Parrinello~\cite{behler2014representing} were the first to introduce descriptors for characterizing atomic local environments combined with a shallow multi-layer perceptron to learn the potential energy of molecules. 
In recent years, deep learning (DL) has demonstrated its powerful ability to learn from raw data without any hand-crafted features in many fields and thus attracted more and more attention.
However, the inherent drawback of deep learning, which requires large amounts of data, has become a bottleneck for its application to more scenarios~\cite{batzner20223}.
To alleviate the dependency on data for DL potentials, recent works have incorporated the inductive bias of symmetry into neural network design, known as geometric deep learning (GDL). 
Symmetry describes the conservation of physical laws, i.e., the unchanged physical properties with any transformations such as translations or rotations.
It allows GDL to be extended to limited data scenarios without any data augmentation.

\begin{figure*}[htbp]
    \centering
    \includegraphics[width=\textwidth]{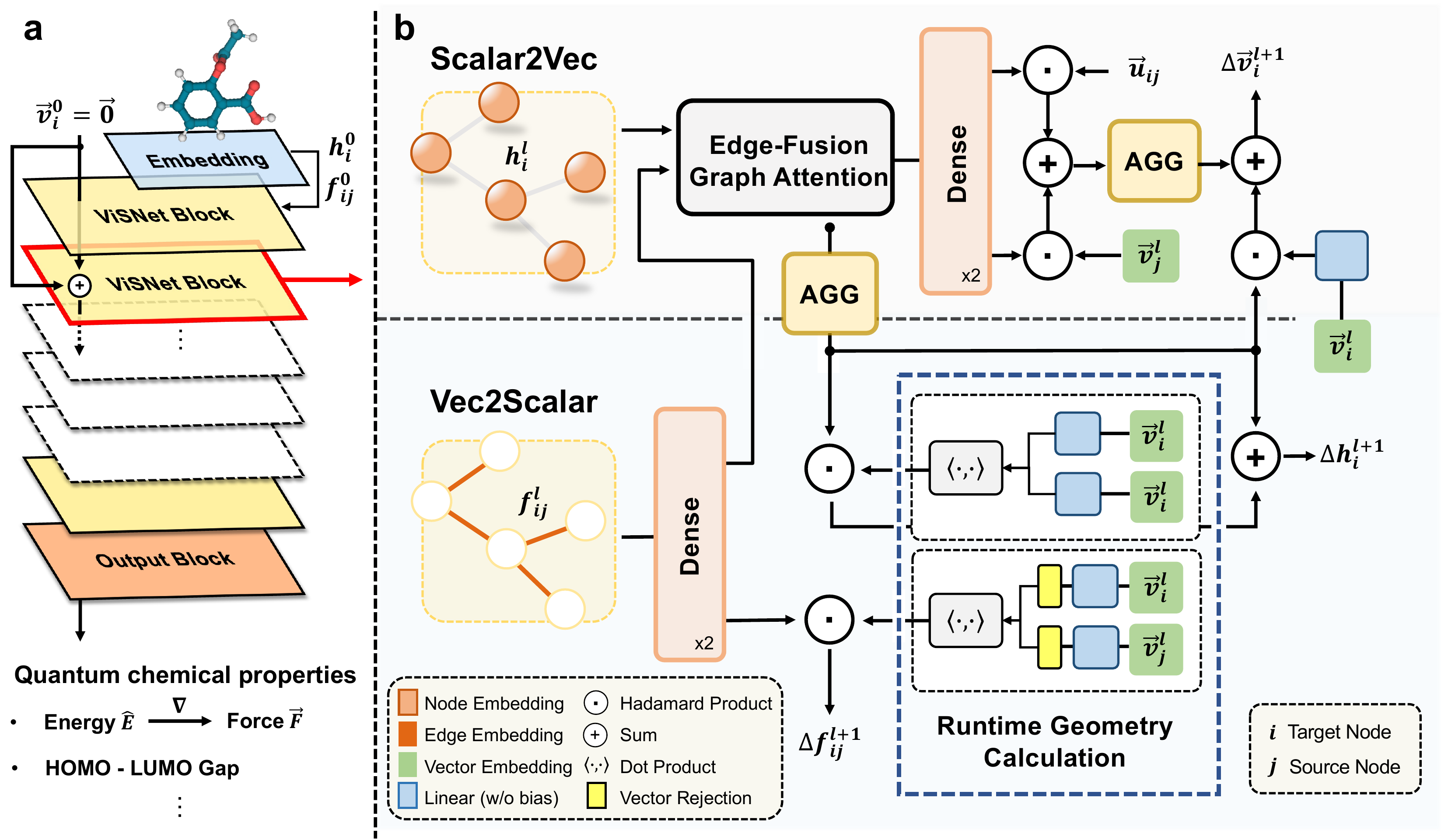}
    \caption{\textbf{The overall architecture of \model.} \textbf{(a)} Model sketch of \model. \model~embeds the 3D structures of molecules and extracts the geometric information through a series of \model~blocks and outputs the molecule properties such as energy, forces, and HOMO-LUMO gap through an output block. \textbf{(b)} Flowchart of one ViSNet Block. One \model~block consists of two modules: i) \textit{Scalar2Vec}, responsible for attaching scalar embeddings to vectors.; ii) \textit{Vec2Scalar}, renovates scalar embeddings built on RGC strategy. The inputs of Scalar2Vec are the node embedding $h_i$, edge embedding $f_{ij}$, direction unit $\Vec v_i$ and the relative positions between two atoms. 
    The edge-fusion graph attention module (serves as $\phi_m^s$) takes as input $h_i$ and the output of the dense layer following $f_{ij}$, and outputs scalar messages. 
    Before aggregation, each scalar message is transformed through a dense layer, then fused with the unit of the relative position $\Vec u_{ij}$ and its own direction unit $\Vec v_j$. 
    We further compute the vector messages and aggregate them all among the neighborhood. 
    Through a gated residual connection, the final residual $\Delta \Vec v_i$ is produced. 
    In Vec2Scalar module, by Hadamard production of aggregated scalar messages and the output of RGC-Angle calculation and adding a gated residual connection, the final $\Delta h_i$ is figured out. Likewise, combining the projected $f_{ij}$ and the output of RGC-Dihedral calculation, the final $\Delta f_{ij}$ is determined.}
    \label{fig:architecture}
\end{figure*}

\begin{figure*}[htbp]
    \centering
    \includegraphics[width=.95\textwidth]{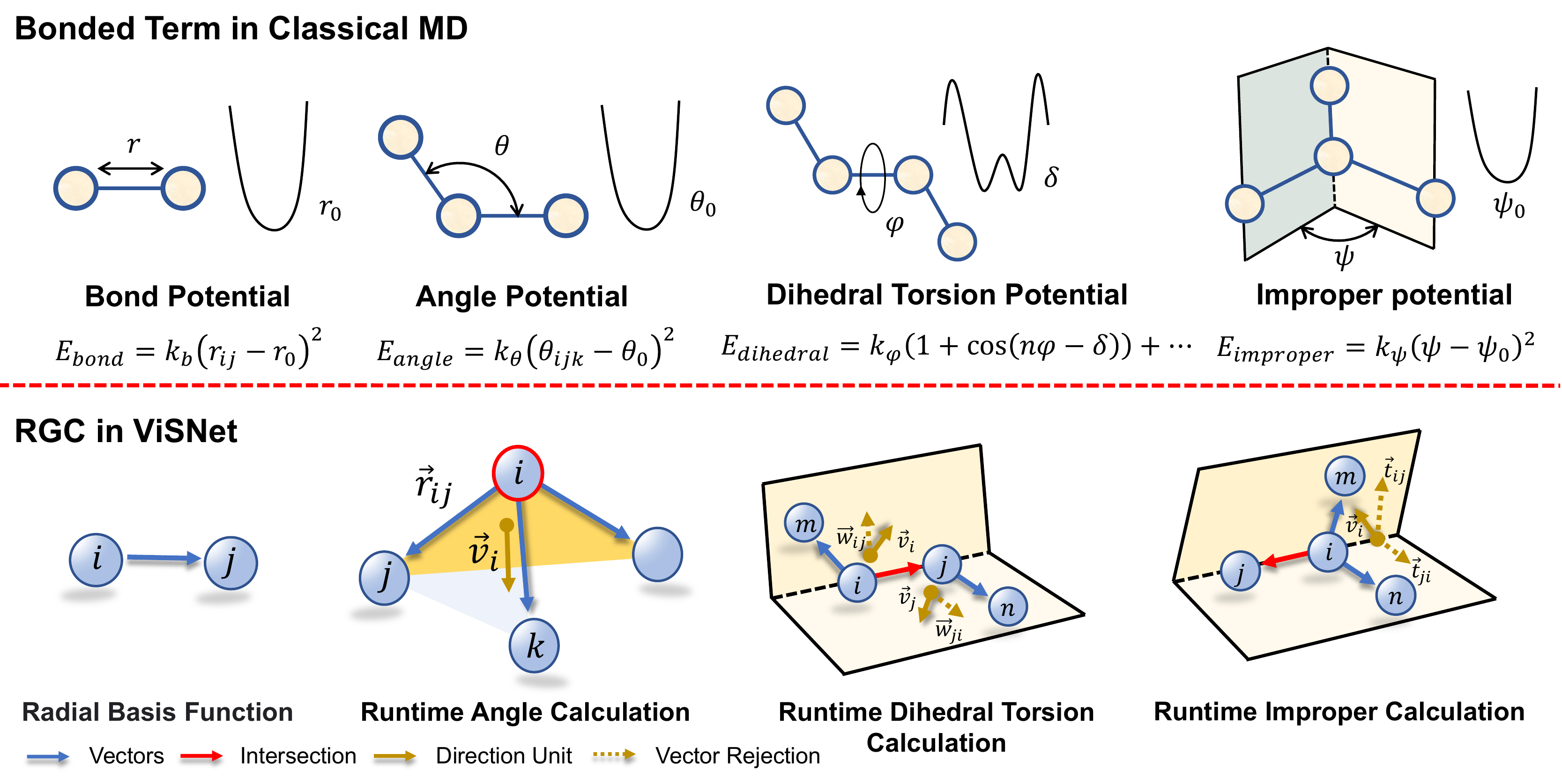}
    \caption{\textbf{Illustration of Runtime Geometry Calculation (RGC) module and its relevance to the potential of bonded terms in classical molecular dynamics.} The bonded terms consist of bond length, bond angle, dihedral torsion and improper angle.  The RGC module depicts all bonded terms of classical MD as model operations in linear time complexity. Yellow arrow $\Vec v_i$ denotes the direction unit in Eq. \ref{unit-sum-vector}.}
    \label{fig:rgc}
\end{figure*}

Equivariant graph neural network (EGNN) is one of the representative approaches in GDL, which has extensive capability to model molecular geometry \cite{batzner20223, brandstetter2021geometric, hutchinson2021lietransformer, fuchs2020se, gasteiger2019directional, gasteiger2020fast, schutt2021equivariant, tholke2022torchmd, gasteiger2021gemnet, unke2021spookynet}. 
A popular kind of EGNN conducts equivariance from directional information and involves geometric features to predict molecular properties.
GemNet~\cite{gasteiger2021gemnet} extends the invariant DimeNet/DimeNet+~\cite{gasteiger2019directional, gasteiger2020fast} with dihedral information.
They explicitly extract geometric information in the Euclidean space with \nth{1}-order geometric tensor, i.e., setting $l_{max}=1$.
PaiNN \cite{schutt2021equivariant} and Equivariant Transformer \cite{tholke2022torchmd} further adopt vector embedding and scalarize the angular representation implicitly via the inner product of the vector embedding itself.
They reduce the complexity of explicit geometry extraction by taking the angular information into consideration.
Another mainstream approach to achieving equivariance is through group representation theory, which can achieve higher accuracy but comes with large computational costs.
NequIP, Allegro, and MACE~\cite{batzner20223, musaelian2022learning, batatia2022mace} achieve state-of-the-art performance on several molecular dynamics simulation datasets leveraging high-order geometric tensors.
On the one hand, algorithms based on group representation theory have strong mathematical foundations and are able to fully utilize geometric information using high-order geometric tensors. 
On the other hand, these algorithms often require computationally expensive operations such as the Clebsch-Gordan product (CG-product)~\cite{han2022geometrically}, making them possibly suitable for periodic systems with elaborate model design but impractical for large molecular systems such as chemical and biological molecules without periodic boundary conditions.

In this study, we propose \model~(short for ``Vector-Scalar interactive graph neural Network"), which alleviates the dilemma between computational costs and sufficient utilization of geometric information.
By incorporating an elaborate Runtime Geometry Calculation (RGC) strategy,  \model~implicitly extracts various geometric features, i.e., angles, dihedral torsion angles, and improper angles in accordance with the force field of classical MD with linear time complexity, thus significantly accelerating model training and inference while reducing the memory consumption.
To extend the vector representation, we introduce spherical harmonics and simplify the computationally expensive Clebsch-Gordan product with the inner product. 
Furthermore, we present a well-designed Vector-Scalar interactive equivariant Message Passing (ViS-MP) mechanism, which fully utilizes the geometric features by interacting vector hidden representations with scalar ones.
When comprehensively evaluated on some benchmark datasets, \model~outperforms all state-of-the-art algorithms on all molecules in MD17, revised MD17 and MD22 datasets and shows superior performance on QM9, Molecule3D dataset indicating the powerful capability of molecular geometric representation. \model~also has won PCQM4Mv2 track in the OGB-LCS@NeurIPS2022 competition (https://ogb.stanford.edu/neurips2022/results/).
We then performed molecular dynamics simulations for each molecule on MD17 driven by \model~trained only with limited data (950 samples). 
The highly consistent interatomic distance distributions and the explored potential energy surfaces between \model~and quantum simulation illustrate that \model~is genuinely data-efficient and can perform simulations with high fidelity.
To further explore the usefulness of \model~to real-world applications, we used an in-house dataset that consists of about 10,000 different conformations of the 166-atom protein Chignolin derived from replica exchange molecular dynamics and calculated at DFT-level. 
When evaluated on the dataset, \model~also achieved significantly better performance than empirical force fields, and the simulations performed by \model~exhibited very close force calculation to DFT.
In addition, \model~exibits reasonable interpretability to map geometric representation to molecular structures.
The contributions of \model~can be summarized as follows:
\begin{itemize}
\item Proposing RGC module that utilizes high-order geometric tensors to implicitly extract various geometric features, including angles, dihedral torsion angles, and improper angles, with linear time complexity.  
\item Introducing ViS-MP mechanism to enable efficient interaction between vector hidden representations and scalar ones and fully exploit the geometric information.
\item Achieving the state-of-the-art performance in 6 benchmarks for predicting energy, forces, HOMO-LUMO gap, and other quantum properties of molecules.
\item Performing molecular dynamics simulations driven by \model~on both small molecules and 166-atom Chignolin with high fidelity.
\item Demonstrating reasonable model interpretability between geometric features and molecular structures.
\end{itemize}

\section{Results}
\subsection{Overview of \model}\label{sec:framework}
\model~is a versatile EGNN which predicts potential energy, atomic forces as well as various quantum chemical properties by taking atomic coordinates and numbers as inputs.
As shown in Fig.\ref{fig:architecture}(a), the model is composed of an embedding block and multiple stacked \model~blocks, followed by an output block. 
The atomic number and coordinates are fed into the embedding block followed by \model~blocks to extract and encode geometric representations. 
The geometric representations are then used to predict molecular properties through the output block. 
It is worth noting that \model~is an energy-conserving potential, i.e., the predicted atomic forces are derived from the negative gradients of the potential energy with respect to the coordinates \cite{chmiela2018towards}.

\noindent
\textbf{RGC: \underline{R}untime \underline{G}eometry \underline{C}alculation}
The success of classical force fields shows that geometric features such as interatomic distances, angles, and dihedral torsioin angles, and improper angles in Fig.\ref{fig:rgc} are essential to determine the total potential energy of molecules.
The explicit extraction of invariant geometric representations in previous studies often suffer from a large amount of time or memory consumption during model training and inference. Given an atom, the calculation of angular information scales $\mathcal{O}(\mathcal{N}^{2})$ with the number of neighboring atoms, while the computational complexity is even $\mathcal{O}(\mathcal{N}^{3})$ for dihedrals \cite{gasteiger2021gemnet}. 
To alleviate this problem, inspired by \cite{schutt2021equivariant}, we propose runtime geometry calculation (RGC), which uses an equivariant vector representation (termed as ``\textit{direction unit}") for each node to preserve its geometric information.
RGC directly calculates the geometric information from the direction unit which only sums the vectors from the target node to its neighbors once. 
Therefore, the computational complexity can be reduced to $\mathcal{O}(\mathcal{N})$.

Considering the sub-structure of a toy molecule with four atoms shown in Fig.~\ref{fig:rgc}, the angular information of the target node $i$ could be conducted from the vector $\Vec r_{ij}$ as follows:
\begin{equation}
\label{unit-sum-vector}
    \Vec u_{ij} = \frac{\Vec r_{ij}}{\left \Vert\Vec r_{ij} \right \Vert}, \quad \Vec v_{i} = \sum_{j=1}^{N_{i}}\Vec u_{ij}
\end{equation}
\begin{equation}
\label{angle-inner}
    {\left \Vert \Vec v_{i} \right \Vert}^{2} = 
    \sum_{j=1}^{N_{i}}\sum_{k=1}^{N_{i}}\left \langle \Vec u_{ij}, \Vec u_{ik} \right \rangle = 
    \sum_{j=1}^{N_{i}}\sum_{k=1}^{N_{i}}\cos \theta_{jik}
\end{equation}
where $\Vec r_{ij}$ is the vector from node $i$ to its neighboring node $j$, $\Vec u_{ij}$ is the unit vector of $\Vec r_{ij}$. 
Here, we define the \textit{direction unit} $\Vec v_{i}$ as the sum of all unit vectors from node $i$ to its all neighboring nodes $j$, where node $i$ is the intersection of all unit vectors. 
As shown in Eq.~\ref{angle-inner}, we calculate the inner product of direction unit $\Vec v_{i}$ which represents the sum of inner products of unit vectors from node $i$ to all its neighboring nodes. 
Combining with Eq.~\ref{unit-sum-vector}, the inner product of direction $\Vec v_{i}$ finally stands for the sum of cosine values of all angles formed by node $i$ and any two of its neighboring nodes.

Similar to runtime angle calculation, we also calculate the vector rejection \cite{perwass2009geometric} of the direction unit $\Vec v_{i}$ of node $i$ and $\Vec v_{j}$ of node $j$ on the vector ${\Vec u_{ij}}$ and ${\Vec u_{ji}}$, respectively. 
\begin{equation}
    \begin{aligned}
        \Vec w_{ij} = {\rm Rej}_{\Vec u_{ij}}(\Vec v_{i}) & = 
    \Vec v_{i} - \left \langle \Vec v_{i}, \Vec u_{ij} \right \rangle \cdot \Vec u_{ij} \\
    & = \sum_{m=1}^{N_{i}}{\rm Rej}_{\Vec u_{ij}}(\Vec u_{im})\\
    \Vec w_{ji} = {\rm Rej}_{\Vec u_{ji}}(\Vec v_{j}) & =
    \Vec v_{j} - \left \langle \Vec v_{j}, \Vec u_{ji} \right \rangle \cdot \Vec u_{ji} \\
    & = \sum_{n=1}^{N_{j}}{\rm Rej}_{\Vec u_{ji}}(\Vec u_{jn})
    \end{aligned}
\end{equation}
where ${\rm Rej}_{\Vec b}(\Vec a)$ represents the vector component of $\Vec a$ perpendicular to $\Vec b$, termed as the vector rejection. $\Vec u_{ij}$ and $\Vec v_{i}$ are defined in Eq.~\ref{unit-sum-vector}. $\Vec w_{ij}$ represents the sum of the vector rejection ${\rm Rej}_{\Vec u_{ij}}(\Vec u_{im})$ and $\Vec w_{ji}$ represents the sum of the vector rejection ${\rm Rej}_{\Vec u_{ji}}(\Vec u_{jn})$.
The inner product between $\Vec w_{ij}$ and $\Vec w_{ji}$ is then calculated to conduct dihedral torsion angle information of the intersecting edge $e_{ij}$ as follows:
\begin{equation}
\label{dihedral}
\begin{aligned}
    \left \langle \Vec w_{ij}, \Vec w_{ji} \right \rangle
        &= \sum_{m=1}^{N_{i}}\sum_{n=1}^{N_{j}}\left \langle {\rm Rej}_{\Vec u_{ij}}(\Vec u_{im}), {\rm Rej}_{\Vec u_{ji}}(\Vec u_{jn}) \right \rangle \\
        &= \sum_{m=1}^{N_{i}}\sum_{n=1}^{N_{j}}\cos \varphi_{mijn}  
\end{aligned}
\end{equation}

The improper angle is derived from a pyramid structure forming by 4 nodes.
As the last toy molecule shown in Fig. \ref{fig:rgc}, node $i$ is the vertex of the pyramid, and the improper torsion angle is formed by two adjacent planes with an intersecting edge $e_{ij}$.
We can also calculate the improper angle by vector rejection:
\begin{equation}
    \begin{aligned}
        \Vec t_{ij} = {\rm Rej}_{\Vec u_{ij}}(\Vec v_{i}) & = \sum_{m=1}^{N_{i}}{\rm Rej}_{\Vec u_{ij}}(\Vec u_{im})\\
        \Vec t_{ji} = {\rm Rej}_{\Vec u_{ji}}(\Vec v_{i}) & =
        \sum_{n=1}^{N_{i}}{\rm Rej}_{\Vec u_{ji}}(\Vec u_{in})
    \end{aligned}
\end{equation}
In the same way, the inner product between $\Vec t_{ij}$ and $\Vec t_{ji}$ indicates the summation of improper angle information formed by $e_{ij}$:
\begin{equation}
\label{improper}
\begin{aligned}
    \left \langle \Vec t_{ij}, \Vec t_{ji} \right \rangle
        &= \sum_{m=1}^{N_{i}}\sum_{n=1}^{N_{i}}\left \langle {\rm Rej}_{\Vec u_{ij}}(\Vec u_{im}), {\rm Rej}_{\Vec u_{ji}}(\Vec u_{in}) \right \rangle \\
        &= \sum_{m=1}^{N_{i}}\sum_{n=1}^{N_{i}}\cos \psi_{mijn}  
\end{aligned}
\end{equation}

Multiple works have shown the effectiveness of high-order geometric tensors for molecular modeling \cite{batzner20223, musaelian2022learning, zitnick2022spherical, liao2022equiformer}.
However, the computational overheads of these approaches are generally expansive due to the CG-product, impeding their further application for large systems.
In this work, we convert the vectors to high-order representation with \textit{spherical harmonics} but discard CG-product with the inner product following the idea of RGC.
We find that the extended high-order geometric tensors can still represent the above angular information in the form of Legendre polynomials according to the \textit{addition theorem}:
\begin{equation}
    \begin{aligned}
        P_{l}(\cos{\theta_{jik}}) & = P_{l}(\Vec u_{ij} \cdot \Vec u_{ik}) \\ & \propto \sum^{l}_{m=-l}Y_{l,m}(\Vec u_{ij})Y_{l,m}^{*}(\Vec u_{ik})
    \end{aligned}
\end{equation}
where the $P_l$ is the Legendre polynomial of degree $l$, $Y_{l,m}$ denotes the spherical harmonics function and $Y_{l,m}^{*}$ denotes its complex conjugation.
We sum the product of different order $l$ to obtain the scalar angular representation, which is the same operation as inner product.
It is worth noting that such an extension doesn't increase the model size and keeps the model architecture unchanged. 

We also provide a proof about the rotational invariance of RGC strategy in the Section \ref{proof}.

\noindent
\textbf{ViS-MP: \underline{V}ector-\underline{S}calar \underline{i}nteractive \underline{M}essage \underline{P}assing}
{In order to make full use of geometric information and enhance the interaction between scalars and vectors, 
we designed an effective vector-scalar interactive message passing mechanism with respect to the intersecting nodes and edges for angles and dihedrals, respectively.
The key operations in ViS-MP are given as follows:
\begin{align}
    m_{i}^l & = \sum_{j \in \mathcal{N}(i)}\phi_{m}^{s}\left(h_{i}^{l}, h_{j}^{l}, f_{ij}^l\right) \label{eq:scalar-msg} \\
    \Vec m_{i}^{l} & =
    \sum_{j \in \mathcal{N}(i)} \phi_{m}^{v}\left(m_{ij}^l, \Vec r_{ij}, \Vec v_{j}^l\right) \label{eq:vec-msg} \\
    h_i^{l+1} & = \phi_{un}^s\left(h_i^l, m^{l}_{i}, \langle \Vec v^{l}_i, \Vec v^{l}_i \rangle\right) \label{eq:angle} \\
    f_{ij}^{l+1} & = \phi_{ue}^{s}\left(f_{ij}^l, \langle {\rm Rej}_{\Vec r_{ij}}(\Vec v^{l}_{i}), {\rm Rej}_{\Vec r_{ji}}(\Vec v^{l}_{j}) \rangle\right) \label{eq:dihedral} \\
    \Vec v_i^{l+1} & = \phi_{un}^{v}\left(\Vec v_i^l, m^{l}_{i}, \Vec m_{i}^{l} \right) \label{eq:vec-embed}
\end{align}
where $h_i$ denotes the scalar embedding of node $i$, $f_{ij}$ stands for the edge feature between node $i$ and node $j$.
$\Vec v_i$ represents the embedding of the direction unit mentioned in RGC.
The superscript of variables indicates the index of the block that the variables belong to.
We omit the improper angle here for brevity. A comprehensive version is depicted in Supplementary.
ViS-MP extends the conventional message passing, aggregation, and update processes with vector-scalar interactions.
Eq. \ref{eq:scalar-msg} and Eq. \ref{eq:vec-msg} depict our message passing and aggregation processes.
To be concrete, scalar messages $m_{ij}$ incorporating scalar embedding $h_j$, $h_i$, and $f_{ij}$ are passed and then aggregated to node $i$ through a message function $\phi_m^s$ (Eq. \ref{eq:scalar-msg}).
Similar operations are applied for vector messages $\Vec m_{i}^{l}$ of node $i$ that incorporates scalar message $m_{ij}$, vector $\Vec r_{ij}$ and vector embedding $\Vec v_j$ (Eq. \ref{eq:vec-msg}).
Eq. \ref{eq:angle} and Eq. \ref{eq:dihedral} demonstrate the update processes.
$h_i$ is updated by the aggregated scalar message output $m_i$ while the inner product of $\Vec v_i$ is updated through an update function $\phi_{un}^s$.
Then $\Vec f_{ij}$ is updated by the inner product of the rejection of the vector embedding $\Vec v_i$ and $\Vec v_j$ through an update function $\phi_{ue}^s$.
Finally, the vector embedding $\Vec v_i$ is updated by both scalar and vector messages through an update function $\phi_{un}^v$.
Notably, the vectors update function, i.e., $\phi^v$ require to be equivariant.
The detailed message and update functions can be found in the Methods section. 
A proof about the equivariance of ViS-MP can be found in Supplementary Methods.

In summary, the geometric features are extracted by inner products in the RGC strategy 
and the scalar and vector embeddings are cyclically updating each other in ViS-MP so as to learn a comprehensive geometric representation from molecular structures.

\begin{table*}[htbp]
\caption{Mean absolute errors (MAE) of energy (kcal/mol) and force (kcal/mol/$\angstrom$) for 7 small organic molecules on MD17 compared with state-of-the-art algorithms. The best one in each category is highlighted in bold. The last column indicates the percentage of improvements compared to the second-best approach, NequIP.}
\begin{threeparttable}
\label{md17-results}
\resizebox{\linewidth}{!}{
\begin{tabular}{llccccccccc|c}
\toprule
Molecule                         &        & SchNet & DimeNet & PaiNN & SpookyNet & ET            & GemNet$^{1}$ & NequIP$^{2}$   & SO3KRATES & ViSNet   & Improvements      \\ \midrule
\multirow{2}{*}{Aspirin}         & energy & 0.37   & 0.204   & 0.167 & 0.151     & 0.123          & -            & 0.131          & 0.139     & \textbf{0.116} & 11.45\% \\
                                 & forces & 1.35   & 0.499   & 0.338 & 0.258     & 0.253          & 0.217        & 0.184          & 0.236     & \textbf{0.155}  & 15.76\% \\ \midrule
\multirow{2}{*}{Ethanol}         & energy & 0.08   & 0.064   & 0.064 & 0.052     & 0.052          & -            & \textbf{0.051} & 0.061     & \textbf{0.051} & 00.00\% \\
                                 & forces & 0.39   & 0.230   & 0.224 & 0.094     & 0.109          & 0.085        & 0.071          & 0.096     & \textbf{0.060} & 15.49\% \\ \midrule
\multirow{2}{*}{Malondialdehyde} & energy & 0.13   & 0.104   & 0.091 & 0.079     & 0.077          & -            & 0.076          & 0.077     & \textbf{0.075} & 01.32\% \\
                                 & forces & 0.66   & 0.383   & 0.319 & 0.167     & 0.169          & 0.155        & 0.129          & 0.147     & \textbf{0.100} & 22.48\% \\ \midrule
\multirow{2}{*}{Naphthalene}     & energy & 0.16   & 0.122   & 0.116 & 0.116     & \textbf{0.085} & -            & 0.113          & 0.115     & \textbf{0.085} & 24.78\% \\
                                 & forces & 0.58   & 0.215   & 0.077 & 0.089     & 0.061          & 0.051        & \textbf{0.039} & 0.074     & \textbf{0.039} & 00.00\% \\ \midrule
\multirow{2}{*}{Salicylic Acid}  & energy & 0.20   & 0.134   & 0.116 & 0.114     & 0.093          & -            & 0.106          & 0.106     & \textbf{0.092} & 13.21\% \\
                                 & forces & 0.85   & 0.374   & 0.195 & 0.180     & 0.129          & 0.125        & 0.090          & 0.145     & \textbf{0.084} & 06.67\% \\ \midrule
\multirow{2}{*}{Toluene}         & energy & 0.12   & 0.102   & 0.095 & 0.094     & \textbf{0.074} & -            & 0.092          & 0.095     & \textbf{0.074} & 19.57\% \\
                                 & forces & 0.57   & 0.216   & 0.094 & 0.087     & 0.067          & 0.060        & 0.046          & 0.073     & \textbf{0.039} & 15.22\% \\ \midrule
\multirow{2}{*}{Uracil}          & energy & 0.14   & 0.115   & 0.106 & 0.105     & \textbf{0.095} & -            & 0.104          & 0.103     & \textbf{0.095} & 08.65\% \\
                                 & forces & 0.56   & 0.301   & 0.139 & 0.119     & 0.095          & 0.097        & 0.076          & 0.111     & \textbf{0.062} & 18.42\% \\ \bottomrule
\end{tabular}}
\begin{tablenotes}
    \item[$1$] The best results are reported among four variants of GemNet.
    \item[$2$] NequIP only shows the results with $l=3$.
\end{tablenotes}
\end{threeparttable}
\end{table*}

\begin{table*}[htbp]
\caption{Mean absolute errors (MAE) of energy (kcal/mol) and force (kcal/mol/$\angstrom$) for 10 small organic molecules on rMD17 compared with state-of-the-art algorithms. The best one in each category is highlighted in bold.}
\begin{threeparttable}
\label{rmd17-results}
\resizebox{\linewidth}{!}{
\begin{tabular}{llcccccccc}
\toprule
Molecule                        &        & UNiTE$^{1}$ & ACE    & GemNet$^{2}$ & NequlP$^{2}$    & BOTNet          & Allegro         & MACE            & ViSNet$^{3}$    \\ \midrule
\multirow{2}{*}{Aspirin}        & energy & 0.055       & 0.141  & -            & 0.0530          & 0.0530          & 0.0530          & 0.0507          & \textbf{0.0445} \\
                                & forces & 0.175       & 0.413  & 0.2191       & 0.1891          & 0.1960          & 0.1683          & 0.1522          & \textbf{0.1520} \\ \midrule
\multirow{2}{*}{Azobenzene}     & energy & 0.025       & 0.083  & -            & 0.0161          & 0.0161          & 0.0277          & 0.0277          & \textbf{0.0156} \\
                                & forces & 0.097       & 0.251  & -            & 0.0669          & 0.0761          & 0.0600          & 0.0692          & \textbf{0.0585} \\ \midrule
\multirow{2}{*}{Benzene}        & energy & 0.002       & 0.0009 & -            & 0.0009          & \textbf{0.0007} & 0.0069          & 0.0092          & \textbf{0.0007} \\
                                & forces & 0.017       & 0.012  & 0.0115       & 0.0069          & 0.0069          & \textbf{0.0046} & 0.0069          & 0.0056          \\ \midrule
\multirow{2}{*}{Ethanol}        & energy & 0.014       & 0.028  & -            & 0.0092          & 0.0092          & 0.0092          & 0.0092          & \textbf{0.0078} \\
                                & forces & 0.085       & 0.168  & 0.083        & 0.0646          & 0.0738          & \textbf{0.0484} & \textbf{0.0484} & 0.0522          \\ \midrule
\multirow{2}{*}{Malonaldehyde}  & energy & 0.025       & 0.039  & -            & 0.0184          & 0.0184          & 0.0138          & 0.0184          & \textbf{0.0132} \\
                                & forces & 0.152       & 0.256  & 0.1522       & 0.1176          & 0.1338          & \textbf{0.0830} & 0.0945          & 0.0893          \\ \midrule
\multirow{2}{*}{Naphthalene}    & energy & 0.011       & 0.021  & -            & \textbf{0.0046} & \textbf{0.0046} & \textbf{0.0046} & 0.0115          & 0.0057          \\
                                & forces & 0.060       & 0.118  & 0.0438       & 0.0300          & 0.0415          & \textbf{0.0208} & 0.0369          & 0.0291          \\ \midrule
\multirow{2}{*}{Paracetamol}    & energy & 0.044       & 0.092  & -            & 0.0323          & 0.0300          & 0.0346          & 0.0300          & \textbf{0.0258} \\
                                & forces & 0.164       & 0.293  & -            & 0.1361          & 0.1338          & 0.1130          & 0.1107          & \textbf{0.1029} \\ \midrule
\multirow{2}{*}{Salicylic acid} & energy & 0.017       & 0.042  & -            & 0.0161          & 0.0184          & 0.0208          & 0.0208          & \textbf{0.0161} \\
                                & forces & 0.088       & 0.214  & 0.1222       & 0.0922          & 0.0992          & \textbf{0.0669} & 0.0715          & 0.0795          \\ \midrule
\multirow{2}{*}{Toluene}        & energy & 0.010       & 0.025  & -            & 0.0069          & 0.0069          & 0.0092          & 0.0115          & \textbf{0.0059} \\
                                & forces & 0.058       & 0.150  & 0.0507       & 0.0369          & 0.0438          & 0.0415          & 0.0346          & \textbf{0.0264} \\ \midrule
\multirow{2}{*}{Uracil}         & energy & 0.013       & 0.025  & -            & 0.0092          & 0.0092          & 0.0138          & 0.0115          & \textbf{0.0069} \\
                                & forces & 0.088       & 0.152  & 0.0876       & 0.0715          & 0.0738          & \textbf{0.0415} & 0.0484          & 0.0495          \\ \bottomrule
\end{tabular}}
\begin{tablenotes}
    \item[$1$] For a fair comparison, the ``direct learning" results without any extra input are compared.
    \item[$2$] The best results are reported among four variants of GemNet and four orders $l \in \{ 0,1,2,3\}$ of NequIP.
    \item[$3$] \model~can achieve better results with longer convergence time.
\end{tablenotes}
\end{threeparttable}
\end{table*}

\subsection{Accurate quantum chemical property predictions}
\label{subsec:property-prediction}
We evaluated \model~on several prevailing benchmark datasets including MD17~\cite{chmiela2018towards, schutt2017quantum, chmiela2017machine}, revised MD17~\cite{christensen2020role}, MD22~\cite{chmiela2022accurate}, QM9~\cite{ramakrishnan2014quantum}, Molecule3D~\cite{xu2021molecule3d} and OGB-LSC PCQM4Mv2~\cite{hu2021ogblsc} for energy, force, and other molecular property prediction.
MD17 consists of the MD trajectories of 7 small organic molecules; the number of conformations in each molecule dataset ranges from 133,700 to 993,237. The dataset rMD17 is a reproduced version of MD17 with higher accuracy.
MD22 is a newly proposed MD trajectories dataset that presents new challenges with respect to larger system sizes (42 to 370 atoms). 
Large molecules such as proteins, lipids, carbohydrates, nucleic acids, and supramolecules are included in MD22.
QM9 consists of 12 kinds of quantum chemical properties of 133,385 small organic molecules with up to 9 heavy atoms.
Molecule3D is a recently proposed dataset including 3,899,647 molecules collected from PubChemQC with their ground-state structures and corresponding properties calculated by DFT. We focus on the prediction of the HOMO-LUMO gap following ComENet~\cite{wang2022comenet}.
OGB-LSC PCQM4Mv2 is a quantum chemistry dataset originally curated under the PubChemQC including DFT-calculated HOMO-LUMO gap of 3,746,619 molecules.
The 3D conformations are provided for 3,378,606 training molecules but not for the validation and test sets.
The training details of \model~on each benchmark are described in the Methods section.

\noindent
\textbf{Energy and force for MD simulation.} 
We compared \model~with the state-of-the-art algorithms, including DimeNet~\cite{gasteiger2019directional}, PaiNN~\cite{schutt2021equivariant}, SpookyNet~\cite{unke2021spookynet}, ET~\cite{tholke2022torchmd}, GemNet~\cite{gasteiger2021gemnet}, UNiTE~\cite{qiao2022informing}, NequIP~\cite{batzner20223}, SO3KRATES~\cite{frank2022so3krates}, Allegro~\cite{musaelian2022learning}, MACE~\cite{batatia2022mace} and so on.
As shown in Table \ref{md17-results} (MD17) and Table \ref{rmd17-results} (rMD17) and Supplementary Table 2 (MD22), it is remarkable that \model~outperformed the compared algorithms for both small (MD17 and rMD17) and large molecules (MD22) with the lowest mean absolute errors (MAE) of predicted energy and forces.
On the one hand, compared with PaiNN, ET and GemNet, \model~incorporated more geometric information and made full use of geometric information in ViS-MP, which contributes to the performance gains. 
On the other hand, compared with NequIP, Allegro, SO3KRATES, MACE et al, \model~testified the effect of introducing spherical harmonics in RGC module.

\noindent
\textbf{Quantum chemical properties.}
As shown in Table \ref{QM9-results}, \model~also achieved the superior performance for chemical property predictions on QM9.
It outperformed the compared algorithms for 9 of 12 chemical properties and achieved comparable results on the remaining properties. 
Elaborated evaluations on Molecule3D confirmed the high prediction accuracy of \model~as shown in Table \ref{m3d-results}. 
\model~achieved 33.6\% and 6.51\% improvements than the second-best for random split and scaffold split, respectively.
Furthermore, \model~ exhibited good portability to other multimodality methods, e.g., Transformer-M \cite{luo2022one} and outperformed other approaches on OGB-LSC PCQM4Mv2 (see Supplementary Fig. 1).
\model~ also achieved the top winners of PCQM4Mv2 track in the OGB-LCS@NeurIPS2022 competition when testing on unseen molecules~\cite{wang2022ensemble} (\url{https://ogb.stanford.edu/neurips2022/results/}).

\begin{table*}[htbp]
  \caption{Mean absolute errors (MAE) of 12 kinds of molecular properties on QM9 compared with state-of-the-art algorithms. The best one in each category is highlighted in bold.}
  \begin{threeparttable}
  \label{QM9-results}
  \centering
  \resizebox{\linewidth}{!}
  {
  \begin{tabular}{llcccccccccc}
    \toprule
    Target                  & Unit                              & SchNet & EGNN  & DimeNet++ & PaiNN  & SphereNet & PaxNet & ET & ComENet & ViSNet  \\
    \midrule
    $\mu$                   & $mD$                              & 33  & 29 & 29.7 & 12 & 24.5 & 10.8 & 11 & 24.5 &\textbf{9.5}   \\
    $\alpha$                & $ma_{0}^{3}$                      & 235 & 71 & 43.5 & 45 & 44.9 & 44.7 & 59 & 45.2 & \textbf{41.1}  \\
    $\epsilon_{HOMO}$       & $meV$                             & 41 & 29 & 24.6 & 27.6 & 22.8 & 22.8 & 20.3 & 23.1 & \textbf{17.3}                    \\
    $\epsilon_{LUMO}$       & $meV$                             & 34 & 25 & 19.5 & 20.4 & 18.9 & 19.2 & 17.5 & 19.8 & \textbf{14.8}                    \\
    $\Delta \epsilon$       & $meV$                             & 63 & 48 & 32.6 & 45.7 & 31.1 & \textbf{31} & 36.1 & 32.4 & 31.7                    \\
    $\langle R^{2} \rangle$ & $ma_{0}^{2}$                      & 73 & 106 & 331 & 66 & 268 & 93 & 33 & 259 & \textbf{29.8}       \\
    $ZPVE$                  & $meV$                             & 1.7 & 1.55 & 1.21 & 1.28 & \textbf{1.12} & 1.17 & 1.84 & 1.2 & 1.56       \\
    $U_0$                   & $meV$                             & 14 & 11 & 6.32 & 5.85 & 6.26 & 5.9 & 6.15 & 6.59 & \textbf{4.23}                    \\
    $U$                     & $meV$                             & 19 & 12 & 6.28 & 5.83 & 6.36 & 5.92 & 6.38 & 6.82 & \textbf{4.25}                    \\
    $H$                     & $meV$                             & 14 & 12 & 6.53 & 5.98 & 6.33 & 6.04 & 6.16 & 6.86 & \textbf{4.52}                    \\
    $G$                     & $meV$                             & 14 & 12 & 7.56 & 7.35 & 7.78 & 7.14 & 7.62 & 7.98 & \textbf{5.86}                      \\
    $C_{v}$                 & $\frac{\rm mcal}{\rm mol \ \rm K}$ & 33 & 31 & 23 & 24 & \textbf{22} & 23.1 & 26 & 24 & 23     \\
    \bottomrule
  \end{tabular}
  }
  \end{threeparttable}
\end{table*}

\begin{table}[ht]
\centering
\caption{Mean absolute errors (MAE) of HOMO-LUMO gap (eV) on Molecule3D test set for both random and scaffold splits compared with state-of-the-art algorithms.\\}
\label{m3d-results}
\resizebox{.3\textwidth}{!}{
\begin{tabular}{lcc}
\toprule
Model       & Random & Scaffold \\ \midrule
GIN-Virtual & 0.1036 & 0.2371   \\
SchNet      & 0.0428 & 0.1511   \\
DimeNet++   & 0.0306 & 0.1214   \\
SphereNet   & 0.0301 & 0.1182   \\
ComENet     & 0.0326 & 0.1273   \\
ViSNet      & \textbf{0.0200} & \textbf{0.1105}   \\ \bottomrule
\end{tabular}
}
\end{table}

\noindent
\textbf{Computational Efficiency}
To evaluate the computational efficiency of our \model, following~\cite{batatia2022mace}, we compare the time latency of \model~with prevailing models in Fig. \ref{fig:time_latency}.
The latency is defined as the time it takes to compute forces on a structure (i.e., the gradient calculation for a set of input coordinates through the whole deep neural network).
As shown in Fig. \ref{fig:time_latency}, ViSNet (L=2) saved 42.8\% time latency compared with MACE (L=2). 
Notably, despite the use of CG-product, Allegro had a significant speed improvement compared to NequIP and BOTNet. However, ViSNet still saved 6.1\%, 4.1\% and 61\% time latency compared to Allegro with L=1, 2 and 3, respectively.

\begin{figure*}[htbp]
    \centering
    \includegraphics[width=0.8\textwidth]{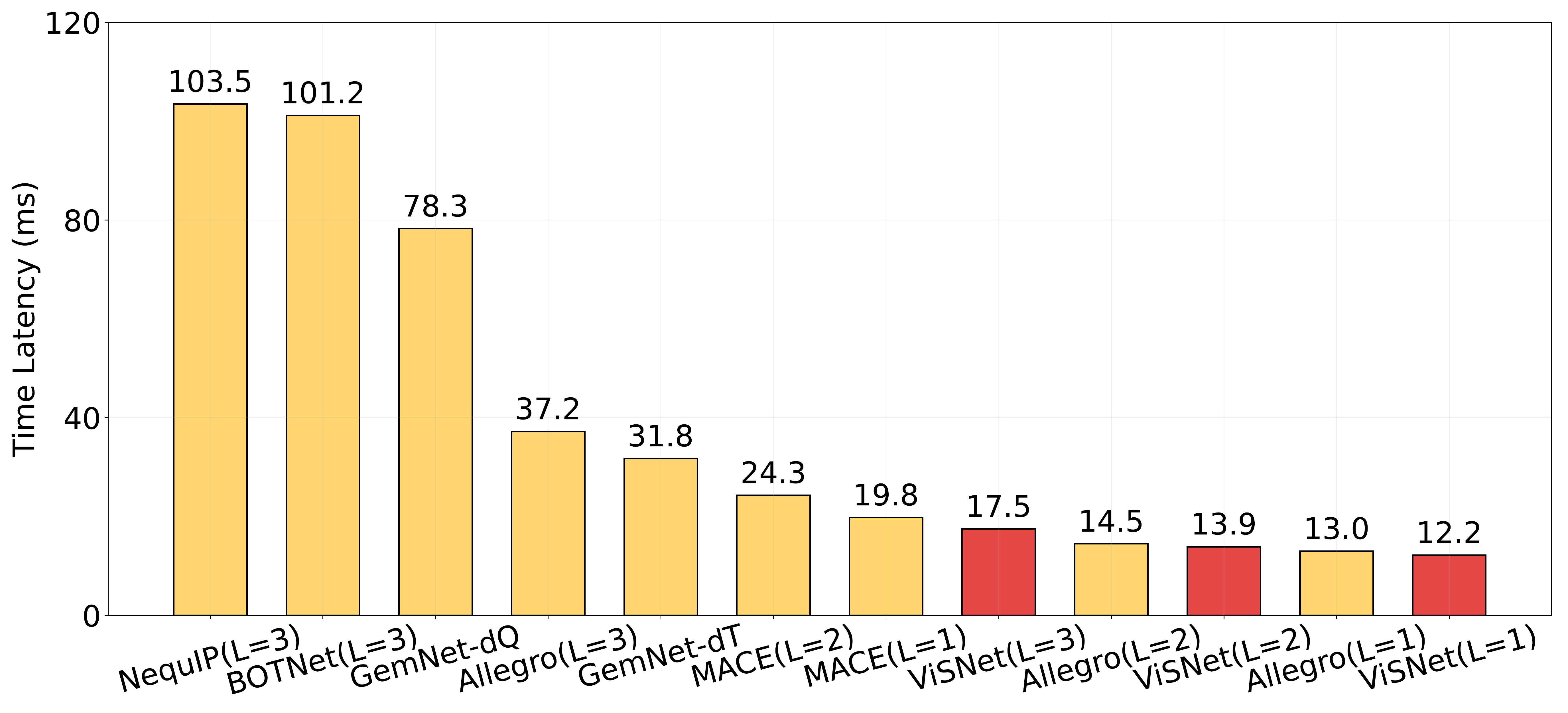}
    \caption{Comparison of time latency with current methods following MACE \cite{batatia2022mace}. Time latency is defined as the time the model takes to compute forces on a structure.
    Experiments are conducted on a Nvidia A100 GPU.}
    \label{fig:time_latency}
\end{figure*}

\subsection{Efficient molecular dynamics simulations on MD17}

\begin{figure*}
    \centering
    \includegraphics[width=\textwidth]{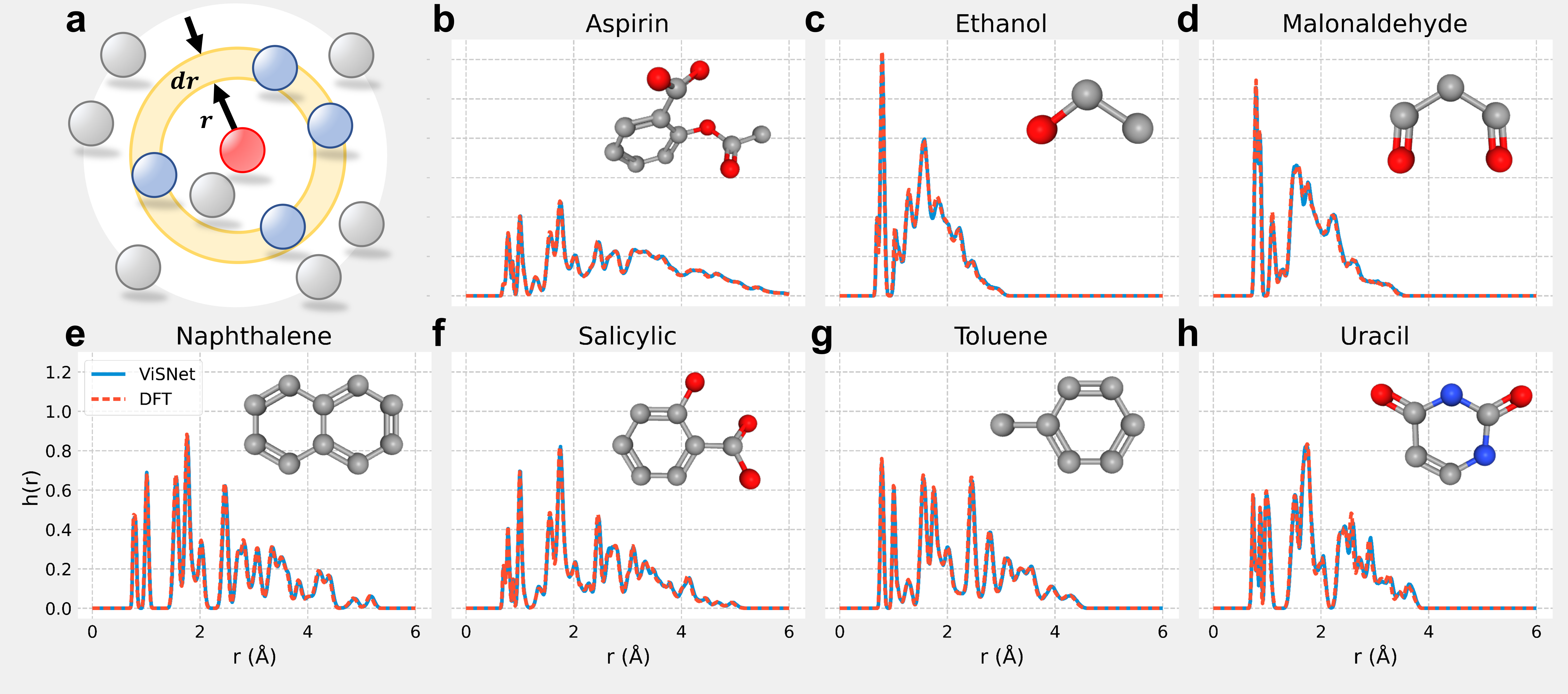}
    \caption{\textbf{The interatomic distance distributions of MD simulations driven by \model~and DFT.} (\textbf{a}) An illustration about the atomic density at a radius $r$ with the arbitrary atom as the center. The interatomic distance distribution is defined as the ensemble average of atomic density. (\textbf{b}) to (\textbf{h}) The interatomic distance distributions comparison between simulations by \model~and DFT for all seven organic molecules in MD17. The curve of \model~is shown using a solid blue line, while the dashed orange line is used for DFT curve. The structures of the corresponding molecules are shown in the upper right corner.}
    \label{fig:rdf}
\end{figure*}

Most of the recently proposed methods are quite accurate in predicting potential energy and atomic forces for the conformations in a given test set. 
Molecular dynamics simulation is one of the important applications of the predicted potential energy and atomic forces.
To evaluate \model~as the potential for molecular dynamics simulation, we incorporated \model~that trained only with 950 samples on MD17 into the ASE simulation framework \cite{larsen2017atomic} to perform MD simulations for all 7 kinds of organic molecules.
All simulations are run with a time step $\tau$ = 0.5 fs under Berendsen thermostat with the other settings the same as those of the MD17 dataset.
As shown in Fig. \ref{fig:rdf},
we analyzed the interatomic distance distributions derived from both AIMD simulations with \model~as the potential and \textit{ab initio} molecular dynamics simulations at DFT level for all 7 molecules, respectively.
As shown in Fig. \ref{fig:rdf}(a), the interatomic distance distribution $h(r)$ is defined as the ensemble average of atomic density at a radius $r$ \cite{chmiela2017machine}.  
Fig. \ref{fig:rdf}(b-h) illustrate the distributions derived from \model~are very close to those generated by DFT.
We also compared the potential energy surfaces sampled by \model~and DFT for these molecules, respectively (Supplementary Fig. 2).
The consistent potential energy surfaces suggest that \model~can well recover the kinetic properties and the conformational space from the simulation trajectories, indicating the usefulness of \model~for real molecular dynamics simulation.
Furthermore, compared with the prohibitive computational cost of DFT, \model~dramatically saves the computational time by 2-3 orders of magnitude (Supplementary Fig. 3 and Supplementary Table 3).
These results demonstrate that with only a few of training samples, \model~can act as the potential to perform high-fidelity molecular dynamics simulations with much less computational cost.

\subsection{Applications for real-world full-atom proteins}
To examine the usefulness of \model~in real-world applications, we made evaluations on the 166-atom protein Chignolin. 
Based on a Chignolin dataset consisting of about 10,000 conformations that sampled by replica exchange MD and calculated at DFT level by Gaussian 16 in our another study,
we split it as training, validation, and test sets by the ratio of 8:1:1. We trained \model~and compared it with molecular mechanics (MM). The DFT results were used as the ground truth.
Fig. \ref{fig:pes}(a) shows the free energy landscape of \textit{Chignolin} and depicted by $d_{Y2-G6}$ (the distance between mainchain O on Y2 and mainchain N on G6) and $d_{E4-T7}$ (the distance between mainchain O on E4 and mainchain N on T7).
The concentrated energy basin on the left shows the folded state and the scattered energy basin on the right shows unfolded state.
We picked six representative structures in the low potential energy regions with both folded and unfolded states and selected some intermediate states with high potential energy colored cyan or blue.
We visualized the energy predictions for the six representative structures, and  
\model~produced a significantly better estimation of the potential energy than MM with empirical force fields did.
Fig. \ref{fig:pes}(b) and (c) show the correlations between the predicted energies by \model~and MM, and the ground truth values calculated by DFT for all conformations in the test set. 
\model~achieved the lower MAE and the higher $R^2$ score.
Similar results can be seen in the force correlations shown in the Supplementary Fig. 4.
Furthermore, we performed MD simulations for Chignolin driven by ViSNet. 10 conformations were randomly selected as initial structures, and 10,000 simulation steps were run for each. As shown in the Fig. \ref{fig:pes}(d), the RMSF for 10 simulation trajectories are shown against simulation steps.
In Fig. \ref{fig:pes}(e), we compared the force differences between ViSNet and those calculated by Gaussian 16 at DFT level. 
The simulation trajectory driven by ViSNet exhibited small force difference to quantum mechanics, which implies that the accuracy and potential usefulness for real-world applications.

\begin{figure*}
    \centering
    \includegraphics[width=\textwidth]{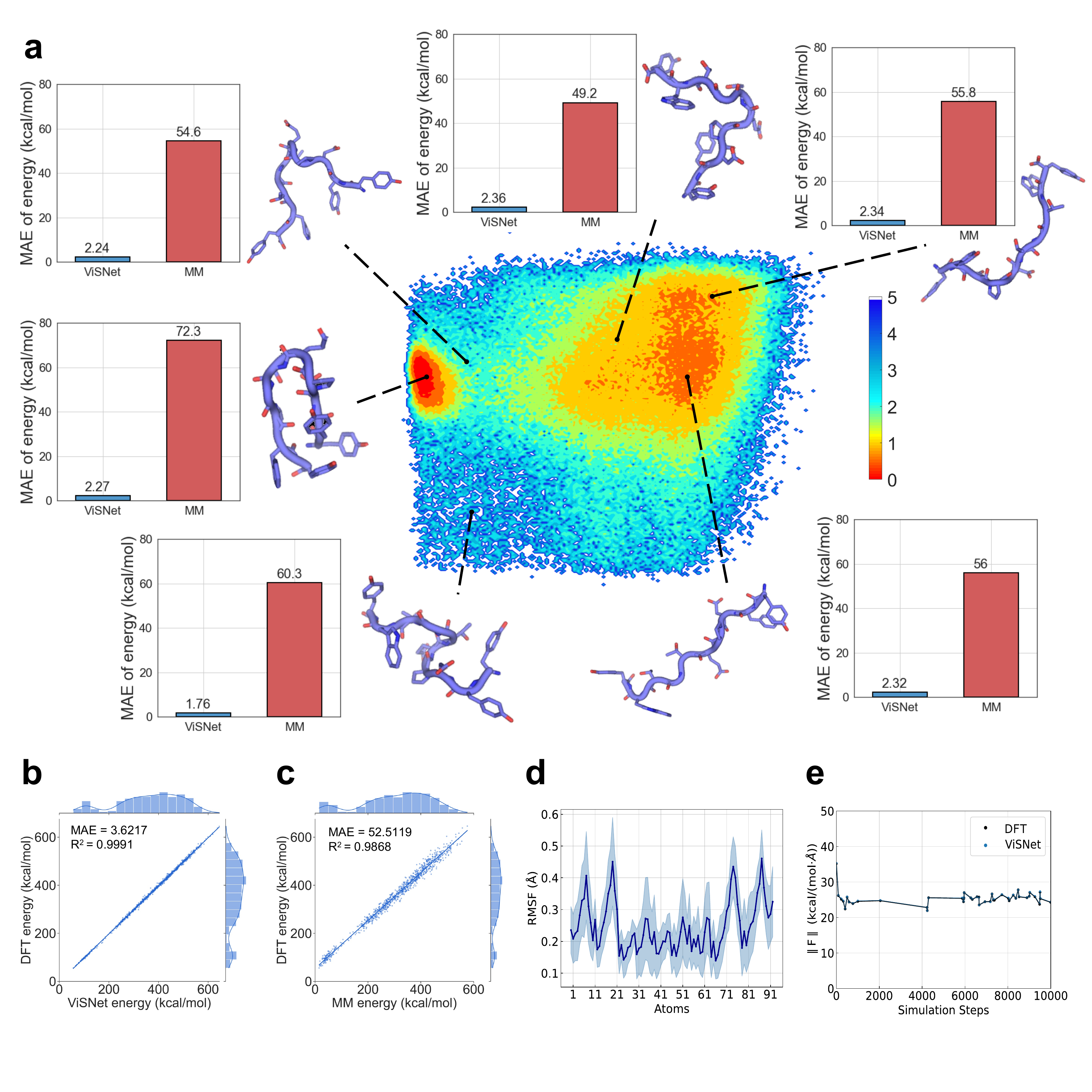}
    \caption{\textbf{Applications of \model~for Chignolin conformational space evaluation and MD simulations.} 
    (\textbf{a}) The energy landscape of \textit{Chignlin} sampled by REMD. The x-axis of the landscape is the distance between mainchain O on Y2 and mainchain N on G6, while the y-axis is the distance between mainchain O on E4 and mainchain N on T7. 6 representative structures were then selected for visualization. Each structure is shown as cartoon and residues are depicted in sticks.
    The histograms show the mean absolute error (MAE) between the energy difference predicted/calculated by \model~or MM, and the ground truth calculated by DFT on the corresponding structure.
    (\textbf{b}) The energy correlations on the test dataset between the ground truth calculated by DFT and the predictions made by \model~. The corresponding distributions of energy predictions or calculations as well as the ground truth are shown.
    (\textbf{c}) The energy correlations on the test dataset between the ground truth calculated by DFT and the predictions made by molecular mechanics.
    (\textbf{d}) The average root mean square fluctuations (RMSF) of the Chignolin trajectories simulated by ViSNet were calculated from 10 different trajectories. The shaded areas indicate the standard deviation range.
    (\textbf{e}) The variation of the force norm during the ViSNet-driven simulation is shown in blue. Multiple frames were randomly selected from the simulation and the ground truth energies and forces were calculated using Gaussian, which are represented by black points.
    }
    \label{fig:pes}
\end{figure*}

\subsection{Interpretability of \model~on molecular structures}
Prior works have shown the effectiveness of incorporating geometric features, such as angles.
The primary method of geometry extraction utilized by ViSNet is the distinct inner product in its runtime geometry calculation.
To this end, we illustrate a reasonable model interpretability of \model~by mapping the angle representations derived from inner product of direction units in the model to the atoms in the molecular structure.
We aim to bridge the gap between geometric representation in \model~and molecular structures.
We visualized the embeddings after the inner product of direction units $\langle \Vec v_i, \Vec v_i \rangle$ extracted from 50 aspirin samples on the validation set. 
The high-dimensional embeddings were reduced to 2-dimensional space using T-SNE \cite{van2008visualizing} and then clustered using DBSCAN \cite{ester1996density} without the prior of the number of clusters.

\begin{figure*}[ht]
    \centering
    \includegraphics[width=.8\textwidth]{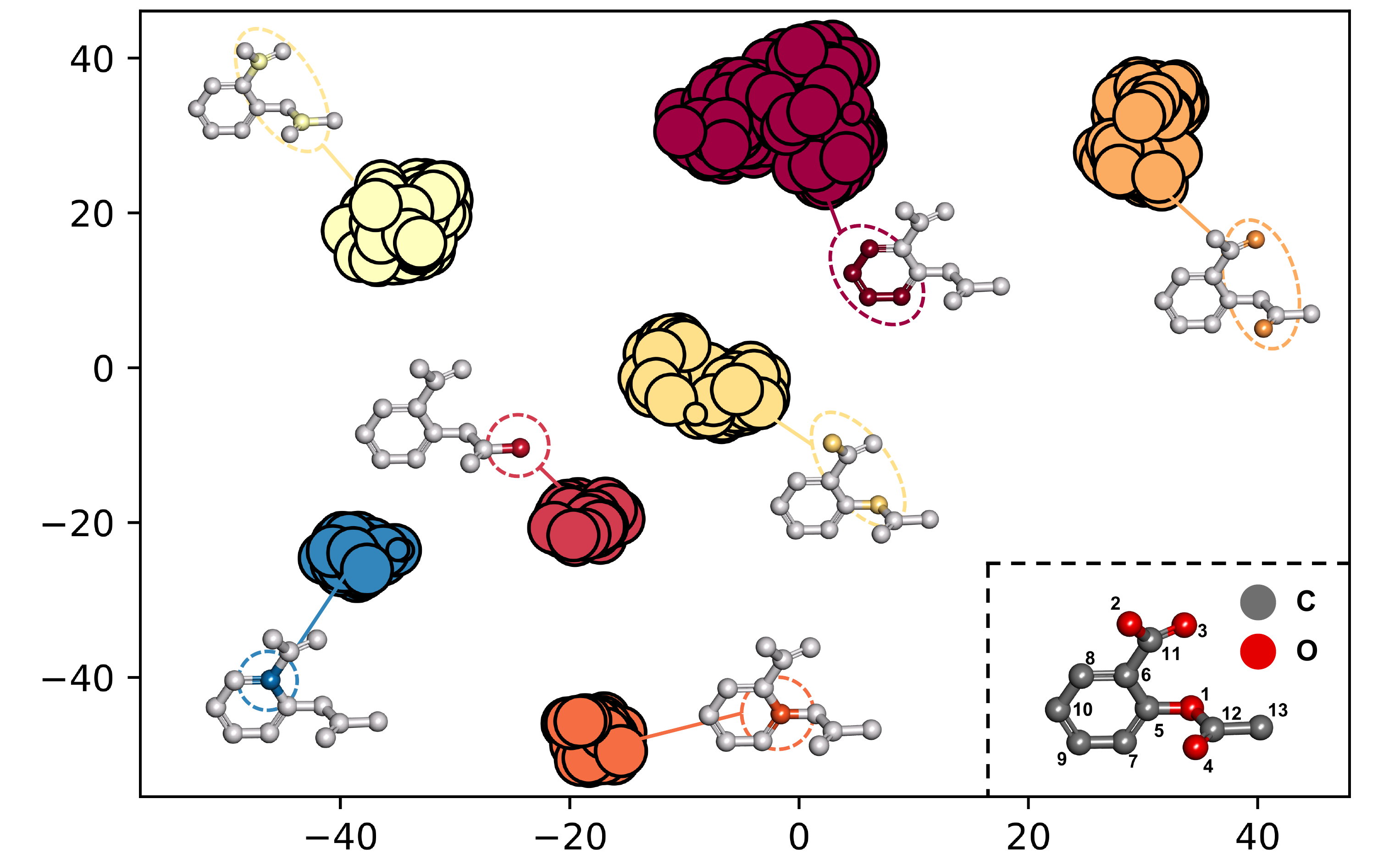}
    \caption{\textbf{Visualization and model interpretability of ViSNet.}
    Clusters of nodes' embeddings after the inner product of the direction units $\langle \Vec v_i, \Vec v_i \rangle$. The $\langle \Vec v_i, \Vec v_i \rangle$ represents angle representations with the intersecting node $i$ as the vertex.
    The atoms in the chemical structure of aspirin corresponding to each cluster are colored with the same color of the cluster, while the remaining atoms are colored light gray.
    A chemical structure of Aspirin and the indices of atoms are illustrated in the bottom right region. Carbon and oxygen atoms are colored dark grey and red, respectively.
    The hydrogen atoms are omitted in both the clustering results and the chemical structure of aspirin for simplification.
    }
    \label{fig:vis}
\end{figure*}

Fig. \ref{fig:vis} exhibits the clustering results of nodes' embeddings after the inner product of their corresponding direction units. We further map the clustered nodes to the atoms of aspirin chemical structure.
Interestingly, the embeddings for these nodes could be distinctly gathered into several clusters shown in different colors. 
For example, although carbon atom $C_{11}$ and carbon atom $C_{12}$ possess different positions and connect with different atoms, their inner product $\left \langle \Vec v_{i}, \Vec v_{i} \right \rangle$ are clustered into the same class for holding similar substructures ($\{C_{11}-O_2O_3C_6\}$ and $\{C_{12}-O_1O_4C_{13}\}$).
To summarize, \model~can discriminate different molecular substructures in the embedding space.

\subsection{Ablation study}
To further explore where the performance gains of \model~come from, we conducted a comprehensive ablation study.
Specifically, we excluded the runtime angle calculation (w/o A), runtime dihedral calculation (w/o D), and both of them (w/o A\&D) in \model, in order to evaluate the usefulness of each part.
ViSNet-improper denotes the additional improper angles and \model$_{l=1}$ uses the \nth{1} order spherical harmonics.

We designed some model variants with different message passing mechanisms based on ViS-MP for scalar and vector interaction.
\model-N directly aggregates the dihedral information to intersecting nodes, and \model-T leverages another form of dihedral calculation. 
The details of these model variants are elaborated in Supplementary. 
The results of the ablation study are shown in Supplementary Table 5 and Supplementary Fig. 5.
Based on the results, we can see that both kinds of directional geometric information are useful and the dihedral information contributes a little bit more to the final performance. The significant performance drop from \model-N and \model-T further validate the effectiveness of ViS-MP mechanism. \model-improper achieves similar performance to \model~for small molecules, but the contribution of improper angles is more obvious for large molecules (see Supplementary Table 2). Furthermore, \model~using higher order spherical harmonics achieves better performance.

\section{Discussion and conclusion}
We propose \model, a novel geometric deep learning potential for molecular dynamics simulation. 
The group representation theory based methods and the directional information based methods are two mainstream classes of geometric deep learning potentials to enforce SE(3) equivariance~\cite{gasteiger2021gemnet}.
\model~takes the advantages from both sides in designing RGC strategy and ViS-MP mechanism. 
On the one hand, the RGC strategy explicitly extracts and exploits the directional geometric information with computationally lightweight operations, making the model training and inference fast. 
On the other hand, ViS-MP employs a series of effective and efficient vector-scalar interactive operations, leading to the full use of the geometric information.    
Furthermore, according to the many-body expansion theory~\cite{nesbet1968atomic, hankins1970water, gordon2012fragmentation}, the potential energy of the whole system equals to the potential of each single atom plus the energy corrections from two-bodies to many-bodies. 
Most of the previous studies model the truncated energy correction terms hierarchically with $k$-hop information via stacking $k$ message passing blocks. 
Different from these approaches, \model~encodes the triplet and quadruplet interactions in a single block, which empowers the model to have much more powerful representation ability. 
In addition, considering that angle and dihedral are important potential terms in empirical force fields, the interpretability of the operations in the RGC strategy provides some insights in constructing hybrid force fields by combining empirical terms with deep learning.

Besides predicting energy, force, and chemical properties with high accuracy, performing molecular dynamics simulations with \emph{ab initio} accuracy at the cost of empirical force field is a grand challenge. 
\model~proves its usefulness in real-world \emph{ab initio} molecular dynamics simulations with less computational costs and the ability of scaling to large molecules such as proteins.
Extending \model~to support larger and more complex molecular systems will be our future research direction.


\section{Methods}

\subsection{Equivariance}
In the context of machine learning for atomic systems, \textit{Equivariance} is a pervasive concept. 
Specifically, the atomic vectors such as dipoles or forces must rotate in a manner consistent with the conformation coordinates.
In molecular dynamics, such equivariance can be ensured by computing gradients based on a predicted conservative scalar energy.
Formally, a function $\mathcal{F}: \mathcal{X} \rightarrow \mathcal{Y}$ is equivariant should guarantee:
\begin{equation}
    \mathcal{F}(\rho_{\mathcal{X}}(g) \circ x) = \rho_{\mathcal{Y}}(g) \circ  \mathcal{F}(x),
\end{equation}
where $\rho_{\mathcal{X}}(g)$ and $\rho_{\mathcal{X}}(g)$ are group representations in input and output spaces.
The integration of equivariance into model parameterization has been shown to be effective, as seen in the implementation of \textit{shift-equivariance} in CNNs, which is critical for enhancing the generalization capacity.
}

\subsection{Proofs of the rotational invariance of RGC}
\label{proof}
Assume that the molecule rotates in 3D space, i.e.,
\begin{equation}
    \Vec{r^{\prime}}_{ij} = R \Vec r_{ij}
\end{equation}
where, $R \in SO(3)$ is an arbitrary rotation matrix that satisfies:
\begin{equation}
    \det \vert R \vert = 1, R^{T}R=I
\end{equation}
The angular information after rotation is calculated as follows:
\begin{equation}
    \Vec{u^{\prime}}_{ij} 
    = \frac{\Vec{r^{\prime}}_{ij}}{\left \Vert\Vec{r^{\prime}}_{ij} \right \Vert}
    = \frac{R \Vec{r}_{ij}}{\det \vert R \vert \cdot \left \Vert\Vec{r}_{ij} \right \Vert}
    = R\Vec{u}_{ij} 
\end{equation}
\begin{equation}
    \Vec{v^{\prime}}_{i} = \sum_{j=1}^{N_{i}}\Vec{u^{'}}_{ij}
    =R\sum_{j=1}^{N_{i}}\Vec{u}_{ij}
    =R\Vec{v}_{i}
\end{equation}
\begin{equation}
\begin{split}
\label{angle-invariance}
    {\left \Vert \Vec{v^{\prime}}_{i} \right \Vert}^{2} = 
    \left \langle \Vec{v^{\prime}}_{i}, \Vec{v^{\prime}}_{i} \right \rangle
    ={(\Vec{v^{\prime}}_{i}})^{T}\Vec{v^{\prime}}_{i} \\
    ={\Vec{v}_{i}}^{T}R^{T}R\Vec{v}_{i}
    =\left \langle \Vec{v}_{i}, \Vec{v}_{i} \right \rangle
    ={\left \Vert \Vec{v}_{i} \right \Vert}^{2}
\end{split}
\end{equation}
As shown in Eq.~\ref{angle-invariance}, the angle information does not change after rotation. The dihedral angular and improper information is also rotationally invariant since:
\begin{equation}
\label{rotate-w}
    \Vec{w^{\prime}}_{ij} = \Vec{v^{\prime}}_{i} - \left \langle \Vec{v^{\prime}}_{i},  \Vec{u^{\prime}}_{ij} \right \rangle \Vec{u^{\prime}}_{ij}
    =R\Vec{v}_{i} - \left \langle R\Vec{v}_{i},  R\Vec{u}_{ij} \right \rangle R\Vec{u}_{ij}
\end{equation}
As Eq.~\ref{angle-invariance} proved, the inner product has rotational invariance. Then, Eq.~\ref{rotate-w} can be further simplified as:
\begin{equation}
    \Vec{w^{\prime}}_{ij}
    = R\left (\Vec{v}_{i} - \left \langle \Vec{v}_{i},  \Vec{u}_{ij} \right \rangle \Vec{u}_{ij} \right)
    = R\Vec{w}_{ij}
\end{equation}
The dihedral or improper angular information after rotation is calculated as:
\begin{equation}
\label{dihedral-invariance}
    \left \langle \Vec{w^{\prime}}_{ij}, \Vec{w^{\prime}}_{ji} \right \rangle = 
    \left \langle R\Vec{w}_{ij}, R\Vec{w}_{ji} \right \rangle = 
    \left \langle \Vec{w}_{ij}, \Vec{w}_{ji} \right \rangle
\end{equation}
As a result, Eq.~\ref{angle-invariance} and Eq.~\ref{dihedral-invariance} have proved the rotational invariance of our proposed runtime geometry calculation (RGC). 

We also provide a proof the equivariance of our ViS-MP in Supplementary Methods.

\subsection{Detailed operations and modules in \model}
\label{methods}
\model~predicts the molecular properties (e.g., energy $\hat{E}$, forces $\Vec F \in \mathbb{R}^{N\times3}$, dipole moment $\mu$) from the current states of atoms, including the atomic positions $X \in \mathbb{R}^{N\times3}$ and atomic numbers $Z \in \mathbb{N}^{N}$. 
The architecture of the proposed \model~is shown in Fig. \ref{fig:architecture}. 
The overall design of \model~follows the vector-scalar interactive message passing as illustrated from Eq. \ref{eq:scalar-msg} - Eq. \ref{eq:dihedral}. 
First, an embedding block encodes the atom numbers and edge distances into the embedding space. 
Then, a series of \model~blocks update the node-wise scalar and vector representations based on their interactions.
A residual connection is placed between two \model~blocks.
Finally, stacked corresponding gated equivariant blocks proposed by \cite{schutt2021equivariant} are attached to the output block for specific molecular property prediction. 

\noindent
\textbf{The Embedding block}
\model~expands the direct node and edge embedding with their neighbors.
It first embeds atomic chemical symbol $z_i$, and calculates the edge representation whose distances within the cutoff through radial basis functions (RBF). 
Then the initial embedding of the atom $i$, its 1-hop neighbors $j$ and the directly connected edge $e_{ij}$ within cutoff are fused together as the initial node embedding $h_i^0$ and edge embedding $f_{ij}^0$. 
In summary, the embedding block is given by:
\begin{equation}
    h_i^0, f_{ij}^0 = \operatorname{Embedding~Block}\left(z_i, z_j, e_{ij}\right), \quad j \in \mathcal{N}(i)
\end{equation}
$\mathcal{N}(i)$ denotes the set of 1-hop neighboring nodes of node $i$, and $j$ is one of its neighbors. 
The embedding process is elaborated in Supplementary. The initial vector embedding $\Vec v_i$ is set to $\Vec 0$.
The vector embeddings $\Vec v$ are projected into the embedding space by following \cite{schutt2021equivariant}; $\Vec v \in \mathbb{R}^{N\times3\times F}$ and $F$ is the size of hidden dimension. 
The advantage of such projection is to assign a unique high-dimensional representation for each embedding to discriminate from each other. Further discussions on its effectiveness and interpretability are given in the Results section.

\noindent
\textbf{The Scalar2Vec module}
In the Scalar2Vec module, the vector embedding $\Vec v$ is updated by both the scalar messages derived from node and edge scalar embeddings (Eq. \ref{eq:scalar-msg}) and the vector messages with inherent geometric information (Eq. \ref{eq:vec-msg}).
The message of each atom is calculated through an Edge-Fusion Graph Attention module, which fuses the node and edge embeddings and computes the attention scores.
The fusion of the node and edge embeddings could be the concatenation operation, Hadamard product, or adding a learnable bias \cite{ying2021transformers}. We leverage the Hadamard product and the \textit{vanilla} multi-head attention mechanism borrowed from Transformer \cite{vaswani2017attention} for edge-node fusion.

Following \cite{tholke2022torchmd}, we pass the fused representations through a nonlinear activation function as shown in Eq. \ref{eq:cutoff}. The value ($V$) in the attention mechanism is also fused by edge features before being multiplied by attention scores weighted by a cosine cutoff as shown in Eq. \ref{eq:value},
\begin{equation}
\label{eq:cutoff}
    \alpha_{ij}^l = \sigma\left( (W_Q^lh_i^l)\left(W_K^lh_j^l\odot\operatorname{Dense}_K^l(f_{ij}^l)\right)^T\right)
\end{equation}
\begin{equation}
\label{eq:value}
    \begin{aligned}
        m_{ij}^l & = \alpha_{ij}^l \cdot \phi(\left \Vert\Vec r_{ij} \right \Vert) \cdot \left(W_V^lh_j^l \odot \operatorname{Dense}_V^l(f_{ij}^l)\right)
    \end{aligned}
\end{equation}
where $l \in \{0,1,2,\cdots, L\}$ is the index of block, $\sigma$ denotes the activation function (SiLU in this paper), $W$ is the learnable weight matrix, $\odot$ represents the Hadamard product, $\phi(\cdot)$ denotes the cosine cutoff and $\operatorname{Dense}(\cdot)$ refers to one learnable weight matrix with activation function. 
For brevity, we omit the learnable bias for linear transformation on scalar embedding in equations, and there is no bias for vector embedding to ensure the equivariance.

Then, the computed $m_{ij}^l$ is used to produce the geometric messages $\Vec m_{ij}^l$ for vectors:
\begin{equation}
    \Vec m_{ij}^l = \left(\operatorname{Dense}_u^l(m_{ij}^l) \odot \Vec u_{ij}\right) + \left(\operatorname{Dense}_v^l(m_{ij}^l) \odot \Vec v_{j}^l\right)
\end{equation}
And the vector embedding $\Vec v^l$ is updated by:
\begin{equation}
    m_i^l = \sum_{j \in \mathcal{N}(i)}m_{ij}^l,  \quad \Vec m_i^l = \sum_{j \in \mathcal{N}(i)} \Vec m_{ij}^l
\end{equation}
\begin{equation}
    \Delta \Vec v_i^{l+1} = \Vec m_i^l + W_{\operatorname{vm}}^l m_i^l \odot W_{\operatorname{v}}^l\Vec v_i^l
\end{equation}

\noindent
\textbf{The Vec2Scalar module}
In the Vec2Scalar module, the node embedding $h_i^l$ and edge embedding $f_{ij}^l$ are updated by the geometric information extracted by the RGC strategy, i.e., angles (Eq. \ref{eq:angle}) and dihedrals (Eq. \ref{eq:dihedral}), respectively.
The residual node embedding $\Delta h_i^{l+1}$, 
is calculated by a Hadamard product between the runtime angle information and the aggregated scalar messages with a gated residual connection:
\begin{equation}
    \Delta h_i^{l+1} = \left\langle W_t^l\Vec v_i^l, W_s^l\Vec v_i^l \right\rangle \odot W_{\operatorname{Angle}}^lm_i^l + W_{\operatorname{res}}^lm_i^l
\end{equation}
To compute the residual edge embedding $\Delta f_{ij}^{l+1}$, we perform the Hadamard product of the runtime dihedral information with the transformed edge embedding:
\begin{equation}
    \begin{aligned}
        \Delta f_{ij}^{l+1} = \left\langle \operatorname{Rej}_{\Vec r_{ij}}(W_{Rt}^l\Vec v_i^l), \operatorname{Rej}_{\Vec r_{ji}}(W_{Rs}^l\Vec v_j^l) \right\rangle \odot \\ \operatorname{Dense}_{\operatorname{Dihedral}}^l(f_{ij}^{l})
    \end{aligned}
\end{equation}
After the residual hidden representations are calculated, we add them to the original input of block $l$ and feed them to the next block.

A comprehensive version which includes improper angles is depicted in Supplementary Methods. 

\noindent
\textbf{The output block}
Following PaiNN \cite{schutt2021equivariant}, we update the scalar embedding and vector embedding of nodes with multiple gated equivariant blocks:
\begin{equation}
    t_i^{l} = \operatorname{Dense}_{o_2}^l([\Vert W_{o_1}^l \Vec v_i^{l} \Vert, h_i^l])
\end{equation}
\begin{equation}
    h_i^{l+1} = W_{o_3}^{l}t_i^{l}
\end{equation}   
\begin{equation}
    \Vec v_i^{l+1} = W_{o_4}^l \Vec v_i^l \odot W_{o_5}^{l}t_i^{l}
\end{equation}
where $[\cdot, \cdot]$ is the tensor concatenation operation.
The final scalar embedding $h_i^L \in \mathbb{R}^{N\times1}$ and vector embedding $\Vec v_i^L \in \mathbb{R}^{N\times3\times 1}$ are used to predict various molecular properties.

On QM9, the molecular dipole is calculated as follows:
\begin{equation}
    \mu = \left \Vert \sum_{i=1}^N \Vec v_i^L + h_i^L(\Vec r_i  - \Vec r_c)\right \Vert
\end{equation}
where $\Vec r_c$ denotes the center of mass. 
Similarly, for the prediction of electronic spatial extent $\langle R^2 \rangle$, we use the following equation:
\begin{equation}
    \langle R^2 \rangle = \sum_{i=1}^N h_i^L \Vert \Vec r_i  - \Vec r_c \Vert^2
\end{equation}
For the remaining 10 properties $y$, we simply aggregate the final scalar embedding of nodes as follows:
\begin{equation}
    y = \sum_{i=1}^N h_i^L
\end{equation}

For models trained on the molecular dynamics datasets including MD17, revised MD17, and \textit{Chignolin}, the total potential energy is obtained as the sum of the
final scalar embedding of the nodes. As an energy-conserving potential, the forces are then calculated using the negative gradients of the predicted total potential energy with respect to the atomic coordinates:
\begin{equation}
    E = \sum_{i=1}^N h_i^L
\end{equation}   
\begin{equation}
    \Vec F_i = - \nabla_{i} E
\end{equation}

\subsection{Dataset splitting schemes}
For the QM9 dataset, we randomly split it into 110,000 samples as the train set, 10,000 samples as the validation set, and the rest as the test set by following the previous studies \cite{schutt2021equivariant,tholke2022torchmd}.
For the Molecule3D and OGB-LSC PCQM4Mv2 dataset, the splitting has been provided in their paper~\cite{xu2021molecule3d, hu2021ogblsc}.

To evaluate the effectiveness of \model~to simulation data, \model~was trained on MD17 and rMD17 with a limited data setting, which consists of only 950 uniformly sampled conformations for model training and 50 conformations for validation for each molecule. For MD22 dataset, we uses the same number of molecules as~\cite{chmiela2022accurate} for training and validation, and the rest as the test set.

Furthermore, the whole \textit{Chignolin} dataset was randomly split into 80\%, 10\%, and 10\% as the training, validation, and test datasets. 
Six representative conformations were picked from the test set for illustration.

\subsection{Experimental settings}
For the QM9 dataset, we adopted a batch size of 32 and a learning rate of 1e-4 for all the properties. For the Molecule3D dataset, we adopted a larger batch size of 512 and a learning rate of 2e-4. For the OGB-LSC PCQM4Mv2 dataset, we trained our model in a mixed 2D/3D mode with a batch size of 256 and a learning rate of 2e-4. The mean squared error (MSE) loss was used for model training.
For the molecular dynamic dataset including MD17, rMD17, MD22 and \textit{Chignolin}, we leveraged a combined MSE loss for energy and force prediction. The weight of energy loss was set to 0.05. The weight of forces loss was set to 0.95. The batch size was was chosen from 2, 4, 8 due to the GPU memory and the learning rate was chosen from 1e-4 to 4e-4 for different molecules. 
The cutoff was set to 5 for small molecules in QM9, MD17, rMD17 and Molecule3D, and changed to 4 for \textit{Chignolin} in order to reduce the number of edges in the molecular graphs. For MD22 dataset, the cutoff of relatively small molecules was set to 5, that of bigger molecules was set to 4. Cutoff was not used in OGB-LSC PCQM4Mv2 dataset.
We used the learning rate decay if the validation loss stopped decreasing. 
The patience was set to 5 epochs for Molecule3D, 15 epochs for QM9, and 30 epochs for MD17, rMD17, MD22 and \textit{Chignolin}.
The learning rate decay factor was set to 0.8 for these models.
We also adopted the early stopping strategy to prevent over-fitting.
The \model~model trained on the molecular dynamic datasets and Molecule3D had 9 hidden layers and the embedding dimension was set to 256. We used a larger model for QM9 dataset, i.e., the embedding dimension changed to 512. For OGB-LSC PCQM4Mv2 dataset, we use the 12-layer and 768-dimension Transformer-M~\cite{luo2022one} as backbone.
More details about the hyperparameters of \model~can be found in Supplementary Table 5.
Experiments were conducted on NVIDIA 32G-V100 GPUs.

\section*{Author contributions}
T. W. led, conceived and designed the study. T. W. is the lead contact.
Y. W., S. L., X. H. and M. L. conducted the work when they were visiting Microsoft Research.
S. L., Y. W. and T. W. carried out algorithm design. Y. W., S. L. and X. H. carried out experiments, evaluations, analysis and visualization. Y. W. and S. L. wrote the original manuscript. T. W., X. H., M. L., Z. W. and B. S revised the manuscript. N. Z. and T. L. contributed to writing. All authors reviewed the final manuscript. 

\section*{Data availability}
The MD17, rMD17, MD22, QM9, Molecule3D and OGB-LSC PCQM4Mv2 dataset are available at their official website. The Chignolin dataset used in this study will be publicly available once the manuscript is published online.

\section*{Code availability}
The code used to produce our results will be publicly available once the manuscript is published online.










\bibliographystyle{naturemag}
\bibliography{mybib}

\begin{thebibliography}{10}
\expandafter\ifx\csname url\endcsname\relax
  \def\url#1{\texttt{#1}}\fi
\expandafter\ifx\csname urlprefix\endcsname\relax\def\urlprefix{URL }\fi
\providecommand{\bibinfo}[2]{#2}
\providecommand{\eprint}[2][]{\url{#2}}

\bibitem{chow20129}
\bibinfo{author}{Chow, E.}, \bibinfo{author}{Klepeis, J.},
  \bibinfo{author}{Rendleman, C.}, \bibinfo{author}{Dror, R.} \&
  \bibinfo{author}{Shaw, D.}
\newblock \bibinfo{title}{9.6 new technologies for molecular dynamics
  simulations}.
\newblock \emph{\bibinfo{journal}{Edward H.. Egelman, editor. Comprehensive
  Biophysics. Amsterdam: Elsevier}} \bibinfo{pages}{86--104}
  (\bibinfo{year}{2012}).

\bibitem{singh2020molecular}
\bibinfo{author}{Singh, S.} \& \bibinfo{author}{Singh, V.~K.}
\newblock \bibinfo{title}{Molecular dynamics simulation: methods and
  application}.
\newblock In \emph{\bibinfo{booktitle}{Frontiers in protein structure,
  function, and dynamics}}, \bibinfo{pages}{213--238}
  (\bibinfo{publisher}{Springer}, \bibinfo{year}{2020}).

\bibitem{lu2021activation}
\bibinfo{author}{Lu, S.} \emph{et~al.}
\newblock \bibinfo{title}{Activation pathway of a g protein-coupled receptor
  uncovers conformational intermediates as targets for allosteric drug design}.
\newblock \emph{\bibinfo{journal}{Nature Communications}}
  \textbf{\bibinfo{volume}{12}}, \bibinfo{pages}{1--15} (\bibinfo{year}{2021}).

\bibitem{li2021exploring}
\bibinfo{author}{Li, Y.} \emph{et~al.}
\newblock \bibinfo{title}{Exploring the regulatory function of the n-terminal
  domain of sars-cov-2 spike protein through molecular dynamics simulation}.
\newblock \emph{\bibinfo{journal}{Advanced theory and simulations}}
  \textbf{\bibinfo{volume}{4}}, \bibinfo{pages}{2100152}
  (\bibinfo{year}{2021}).

\bibitem{kohn1965self}
\bibinfo{author}{Kohn, W.} \& \bibinfo{author}{Sham, L.~J.}
\newblock \bibinfo{title}{Self-consistent equations including exchange and
  correlation effects}.
\newblock \emph{\bibinfo{journal}{Physical review}}
  \textbf{\bibinfo{volume}{140}}, \bibinfo{pages}{A1133}
  (\bibinfo{year}{1965}).

\bibitem{marx2009ab}
\bibinfo{author}{Marx, D.} \& \bibinfo{author}{Hutter, J.}
\newblock \emph{\bibinfo{title}{Ab initio molecular dynamics: basic theory and
  advanced methods}} (\bibinfo{publisher}{Cambridge University Press},
  \bibinfo{year}{2009}).

\bibitem{christensen2020fchl}
\bibinfo{author}{Christensen, A.~S.}, \bibinfo{author}{Bratholm, L.~A.},
  \bibinfo{author}{Faber, F.~A.} \& \bibinfo{author}{Anatole~von Lilienfeld,
  O.}
\newblock \bibinfo{title}{Fchl revisited: Faster and more accurate quantum
  machine learning}.
\newblock \emph{\bibinfo{journal}{The Journal of chemical physics}}
  \textbf{\bibinfo{volume}{152}}, \bibinfo{pages}{044107}
  (\bibinfo{year}{2020}).

\bibitem{bartok2010gaussian}
\bibinfo{author}{Bart{\'o}k, A.~P.}, \bibinfo{author}{Payne, M.~C.},
  \bibinfo{author}{Kondor, R.} \& \bibinfo{author}{Cs{\'a}nyi, G.}
\newblock \bibinfo{title}{Gaussian approximation potentials: The accuracy of
  quantum mechanics, without the electrons}.
\newblock \emph{\bibinfo{journal}{Physical review letters}}
  \textbf{\bibinfo{volume}{104}}, \bibinfo{pages}{136403}
  (\bibinfo{year}{2010}).

\bibitem{behler2014representing}
\bibinfo{author}{Behler, J.}
\newblock \bibinfo{title}{Representing potential energy surfaces by
  high-dimensional neural network potentials}.
\newblock \emph{\bibinfo{journal}{Journal of Physics: Condensed Matter}}
  \textbf{\bibinfo{volume}{26}}, \bibinfo{pages}{183001}
  (\bibinfo{year}{2014}).

\bibitem{batzner20223}
\bibinfo{author}{Batzner, S.} \emph{et~al.}
\newblock \bibinfo{title}{E (3)-equivariant graph neural networks for
  data-efficient and accurate interatomic potentials}.
\newblock \emph{\bibinfo{journal}{Nature communications}}
  \textbf{\bibinfo{volume}{13}}, \bibinfo{pages}{1--11} (\bibinfo{year}{2022}).

\bibitem{brandstetter2021geometric}
\bibinfo{author}{Brandstetter, J.}, \bibinfo{author}{Hesselink, R.},
  \bibinfo{author}{van~der Pol, E.}, \bibinfo{author}{Bekkers, E.} \&
  \bibinfo{author}{Welling, M.}
\newblock \bibinfo{title}{Geometric and physical quantities improve e (3)
  equivariant message passing}.
\newblock \emph{\bibinfo{journal}{International Conference on Learning
  Representations}}  (\bibinfo{year}{2022}).

\bibitem{hutchinson2021lietransformer}
\bibinfo{author}{Hutchinson, M.~J.} \emph{et~al.}
\newblock \bibinfo{title}{Lietransformer: Equivariant self-attention for lie
  groups}.
\newblock In \emph{\bibinfo{booktitle}{International Conference on Machine
  Learning}}, \bibinfo{pages}{4533--4543} (\bibinfo{organization}{PMLR},
  \bibinfo{year}{2021}).

\bibitem{fuchs2020se}
\bibinfo{author}{Fuchs, F.}, \bibinfo{author}{Worrall, D.},
  \bibinfo{author}{Fischer, V.} \& \bibinfo{author}{Welling, M.}
\newblock \bibinfo{title}{Se (3)-transformers: 3d roto-translation equivariant
  attention networks}.
\newblock \emph{\bibinfo{journal}{Advances in Neural Information Processing
  Systems}} \textbf{\bibinfo{volume}{33}}, \bibinfo{pages}{1970--1981}
  (\bibinfo{year}{2020}).

\bibitem{gasteiger2019directional}
\bibinfo{author}{Gasteiger, J.}, \bibinfo{author}{Gro{\ss}, J.} \&
  \bibinfo{author}{G{\"u}nnemann, S.}
\newblock \bibinfo{title}{Directional message passing for molecular graphs}.
\newblock In \emph{\bibinfo{booktitle}{International Conference on Learning
  Representations}} (\bibinfo{year}{2019}).

\bibitem{gasteiger2020fast}
\bibinfo{author}{Gasteiger, J.}, \bibinfo{author}{Giri, S.},
  \bibinfo{author}{Margraf, J.~T.} \& \bibinfo{author}{G{\"u}nnemann, S.}
\newblock \bibinfo{title}{Fast and uncertainty-aware directional message
  passing for non-equilibrium molecules}.
\newblock \emph{\bibinfo{journal}{Advances in Neural Information Processing
  Systems}}  (\bibinfo{year}{2020}).

\bibitem{schutt2021equivariant}
\bibinfo{author}{Sch{\"u}tt, K.}, \bibinfo{author}{Unke, O.} \&
  \bibinfo{author}{Gastegger, M.}
\newblock \bibinfo{title}{Equivariant message passing for the prediction of
  tensorial properties and molecular spectra}.
\newblock In \emph{\bibinfo{booktitle}{International Conference on Machine
  Learning}}, \bibinfo{pages}{9377--9388} (\bibinfo{organization}{PMLR},
  \bibinfo{year}{2021}).

\bibitem{tholke2022torchmd}
\bibinfo{author}{Th{\"o}lke, P.} \& \bibinfo{author}{De~Fabritiis, G.}
\newblock \bibinfo{title}{Torchmd-net: Equivariant transformers for neural
  network based molecular potentials}.
\newblock \emph{\bibinfo{journal}{The International Conference on Learning
  Representations}}  (\bibinfo{year}{2022}).

\bibitem{gasteiger2021gemnet}
\bibinfo{author}{Gasteiger, J.}, \bibinfo{author}{Becker, F.} \&
  \bibinfo{author}{G{\"u}nnemann, S.}
\newblock \bibinfo{title}{Gemnet: Universal directional graph neural networks
  for molecules}.
\newblock \emph{\bibinfo{journal}{Advances in Neural Information Processing
  Systems}} \textbf{\bibinfo{volume}{34}}, \bibinfo{pages}{6790--6802}
  (\bibinfo{year}{2021}).

\bibitem{unke2021spookynet}
\bibinfo{author}{Unke, O.~T.} \emph{et~al.}
\newblock \bibinfo{title}{Spookynet: Learning force fields with electronic
  degrees of freedom and nonlocal effects}.
\newblock \emph{\bibinfo{journal}{Nature communications}}
  \textbf{\bibinfo{volume}{12}}, \bibinfo{pages}{1--14} (\bibinfo{year}{2021}).

\bibitem{musaelian2022learning}
\bibinfo{author}{Musaelian, A.} \emph{et~al.}
\newblock \bibinfo{title}{Learning local equivariant representations for
  large-scale atomistic dynamics}.
\newblock \emph{\bibinfo{journal}{arXiv preprint arXiv:2204.05249}}
  (\bibinfo{year}{2022}).

\bibitem{batatia2022mace}
\bibinfo{author}{Batatia, I.}, \bibinfo{author}{Kov{\'a}cs, D.~P.},
  \bibinfo{author}{Simm, G.~N.}, \bibinfo{author}{Ortner, C.} \&
  \bibinfo{author}{Cs{\'a}nyi, G.}
\newblock \bibinfo{title}{Mace: Higher order equivariant message passing neural
  networks for fast and accurate force fields}.
\newblock \emph{\bibinfo{journal}{arXiv preprint arXiv:2206.07697}}
  (\bibinfo{year}{2022}).

\bibitem{han2022geometrically}
\bibinfo{author}{Han, J.}, \bibinfo{author}{Rong, Y.}, \bibinfo{author}{Xu, T.}
  \& \bibinfo{author}{Huang, W.}
\newblock \bibinfo{title}{Geometrically equivariant graph neural networks: A
  survey}.
\newblock \emph{\bibinfo{journal}{arXiv preprint arXiv:2202.07230}}
  (\bibinfo{year}{2022}).

\bibitem{chmiela2018towards}
\bibinfo{author}{Chmiela, S.}, \bibinfo{author}{Sauceda, H.~E.},
  \bibinfo{author}{M{\"u}ller, K.-R.} \& \bibinfo{author}{Tkatchenko, A.}
\newblock \bibinfo{title}{Towards exact molecular dynamics simulations with
  machine-learned force fields}.
\newblock \emph{\bibinfo{journal}{Nature communications}}
  \textbf{\bibinfo{volume}{9}}, \bibinfo{pages}{1--10} (\bibinfo{year}{2018}).

\bibitem{perwass2009geometric}
\bibinfo{author}{Perwass, C.}, \bibinfo{author}{Edelsbrunner, H.},
  \bibinfo{author}{Kobbelt, L.} \& \bibinfo{author}{Polthier, K.}
\newblock \emph{\bibinfo{title}{Geometric algebra with applications in
  engineering}}, vol.~\bibinfo{volume}{4} (\bibinfo{publisher}{Springer},
  \bibinfo{year}{2009}).

\bibitem{zitnick2022spherical}
\bibinfo{author}{Zitnick, C.~L.} \emph{et~al.}
\newblock \bibinfo{title}{Spherical channels for modeling atomic interactions}.
\newblock \emph{\bibinfo{journal}{arXiv preprint arXiv:2206.14331}}
  (\bibinfo{year}{2022}).

\bibitem{liao2022equiformer}
\bibinfo{author}{Liao, Y.-L.} \& \bibinfo{author}{Smidt, T.}
\newblock \bibinfo{title}{Equiformer: Equivariant graph attention transformer
  for 3d atomistic graphs}.
\newblock \emph{\bibinfo{journal}{arXiv preprint arXiv:2206.11990}}
  (\bibinfo{year}{2022}).

\bibitem{schutt2017quantum}
\bibinfo{author}{Sch{\"u}tt, K.~T.}, \bibinfo{author}{Arbabzadah, F.},
  \bibinfo{author}{Chmiela, S.}, \bibinfo{author}{M{\"u}ller, K.~R.} \&
  \bibinfo{author}{Tkatchenko, A.}
\newblock \bibinfo{title}{Quantum-chemical insights from deep tensor neural
  networks}.
\newblock \emph{\bibinfo{journal}{Nature communications}}
  \textbf{\bibinfo{volume}{8}}, \bibinfo{pages}{1--8} (\bibinfo{year}{2017}).

\bibitem{chmiela2017machine}
\bibinfo{author}{Chmiela, S.} \emph{et~al.}
\newblock \bibinfo{title}{Machine learning of accurate energy-conserving
  molecular force fields}.
\newblock \emph{\bibinfo{journal}{Science advances}}
  \textbf{\bibinfo{volume}{3}}, \bibinfo{pages}{e1603015}
  (\bibinfo{year}{2017}).

\bibitem{christensen2020role}
\bibinfo{author}{Christensen, A.~S.} \& \bibinfo{author}{Von~Lilienfeld, O.~A.}
\newblock \bibinfo{title}{On the role of gradients for machine learning of
  molecular energies and forces}.
\newblock \emph{\bibinfo{journal}{Machine Learning: Science and Technology}}
  \textbf{\bibinfo{volume}{1}}, \bibinfo{pages}{045018} (\bibinfo{year}{2020}).

\bibitem{chmiela2022accurate}
\bibinfo{author}{Chmiela, S.} \emph{et~al.}
\newblock \bibinfo{title}{Accurate global machine learning force fields for
  molecules with hundreds of atoms}.
\newblock \emph{\bibinfo{journal}{arXiv preprint arXiv:2209.14865}}
  (\bibinfo{year}{2022}).

\bibitem{ramakrishnan2014quantum}
\bibinfo{author}{Ramakrishnan, R.}, \bibinfo{author}{Dral, P.~O.},
  \bibinfo{author}{Rupp, M.} \& \bibinfo{author}{Von~Lilienfeld, O.~A.}
\newblock \bibinfo{title}{Quantum chemistry structures and properties of 134
  kilo molecules}.
\newblock \emph{\bibinfo{journal}{Scientific data}}
  \textbf{\bibinfo{volume}{1}}, \bibinfo{pages}{1--7} (\bibinfo{year}{2014}).

\bibitem{xu2021molecule3d}
\bibinfo{author}{Xu, Z.} \emph{et~al.}
\newblock \bibinfo{title}{Molecule3d: A benchmark for predicting 3d geometries
  from molecular graphs}.
\newblock \emph{\bibinfo{journal}{arXiv preprint arXiv:2110.01717}}
  (\bibinfo{year}{2021}).

\bibitem{hu2021ogblsc}
\bibinfo{author}{Hu, W.} \emph{et~al.}
\newblock \bibinfo{title}{{OGB}-{LSC}: A large-scale challenge for machine
  learning on graphs}.
\newblock In \emph{\bibinfo{booktitle}{Thirty-fifth Conference on Neural
  Information Processing Systems Datasets and Benchmarks Track (Round 2)}}
  (\bibinfo{year}{2021}).
\newblock \urlprefix\url{https://openreview.net/forum?id=qkcLxoC52kL}.

\bibitem{wang2022comenet}
\bibinfo{author}{Wang, L.}, \bibinfo{author}{Liu, Y.}, \bibinfo{author}{Lin,
  Y.}, \bibinfo{author}{Liu, H.} \& \bibinfo{author}{Ji, S.}
\newblock \bibinfo{title}{Comenet: Towards complete and efficient message
  passing for 3d molecular graphs}.
\newblock \emph{\bibinfo{journal}{Advances in Neural Information Processing
  Systems}}  (\bibinfo{year}{2022}).

\bibitem{qiao2022informing}
\bibinfo{author}{Qiao, Z.} \emph{et~al.}
\newblock \bibinfo{title}{Informing geometric deep learning with electronic
  interactions to accelerate quantum chemistry}.
\newblock \emph{\bibinfo{journal}{Proceedings of the National Academy of
  Sciences}} \textbf{\bibinfo{volume}{119}}, \bibinfo{pages}{e2205221119}
  (\bibinfo{year}{2022}).

\bibitem{frank2022so3krates}
\bibinfo{author}{Frank, T.}, \bibinfo{author}{Unke, O.~T.} \&
  \bibinfo{author}{Muller, K.~R.}
\newblock \bibinfo{title}{So3krates: Equivariant attention for interactions on
  arbitrary length-scales in molecular systems}.
\newblock In \emph{\bibinfo{booktitle}{Advances in Neural Information
  Processing Systems}} (\bibinfo{year}{2022}).

\bibitem{luo2022one}
\bibinfo{author}{Luo, S.} \emph{et~al.}
\newblock \bibinfo{title}{One transformer can understand both 2d \& 3d
  molecular data}.
\newblock \emph{\bibinfo{journal}{arXiv preprint arXiv:2210.01765}}
  (\bibinfo{year}{2022}).

\bibitem{wang2022ensemble}
\bibinfo{author}{Wang, Y.} \emph{et~al.}
\newblock \bibinfo{title}{An ensemble of visnet, transformer-m, and pretraining
  models for molecular property prediction in ogb large-scale challenge@
  neurips 2022}.
\newblock \emph{\bibinfo{journal}{arXiv preprint arXiv:2211.12791}}
  (\bibinfo{year}{2022}).

\bibitem{larsen2017atomic}
\bibinfo{author}{Larsen, A.~H.} \emph{et~al.}
\newblock \bibinfo{title}{The atomic simulation environment—a python library
  for working with atoms}.
\newblock \emph{\bibinfo{journal}{Journal of Physics: Condensed Matter}}
  \textbf{\bibinfo{volume}{29}}, \bibinfo{pages}{273002}
  (\bibinfo{year}{2017}).

\bibitem{van2008visualizing}
\bibinfo{author}{Van~der Maaten, L.} \& \bibinfo{author}{Hinton, G.}
\newblock \bibinfo{title}{Visualizing data using t-sne.}
\newblock \emph{\bibinfo{journal}{Journal of machine learning research}}
  \textbf{\bibinfo{volume}{9}} (\bibinfo{year}{2008}).

\bibitem{ester1996density}
\bibinfo{author}{Ester, M.}, \bibinfo{author}{Kriegel, H.-P.},
  \bibinfo{author}{Sander, J.}, \bibinfo{author}{Xu, X.} \emph{et~al.}
\newblock \bibinfo{title}{A density-based algorithm for discovering clusters in
  large spatial databases with noise.}
\newblock In \emph{\bibinfo{booktitle}{kdd}}, vol.~\bibinfo{volume}{96},
  \bibinfo{pages}{226--231} (\bibinfo{year}{1996}).

\bibitem{nesbet1968atomic}
\bibinfo{author}{Nesbet, R.}
\newblock \bibinfo{title}{Atomic {B}ethe-{G}oldstone equations. {III}.
  correlation energies of ground states of {B}e, {B}, {C}, {N}, {O}, {F}, and
  {N}e}.
\newblock \emph{\bibinfo{journal}{Physical Review}}
  \textbf{\bibinfo{volume}{175}}, \bibinfo{pages}{2} (\bibinfo{year}{1968}).

\bibitem{hankins1970water}
\bibinfo{author}{Hankins, D.}, \bibinfo{author}{Moskowitz, J.} \&
  \bibinfo{author}{Stillinger, F.}
\newblock \bibinfo{title}{Water molecule interactions}.
\newblock \emph{\bibinfo{journal}{The Journal of Chemical Physics}}
  \textbf{\bibinfo{volume}{53}}, \bibinfo{pages}{4544--4554}
  (\bibinfo{year}{1970}).

\bibitem{gordon2012fragmentation}
\bibinfo{author}{Gordon, M.~S.}, \bibinfo{author}{Fedorov, D.~G.},
  \bibinfo{author}{Pruitt, S.~R.} \& \bibinfo{author}{Slipchenko, L.~V.}
\newblock \bibinfo{title}{Fragmentation methods: A route to accurate
  calculations on large systems}.
\newblock \emph{\bibinfo{journal}{Chemical reviews}}
  \textbf{\bibinfo{volume}{112}}, \bibinfo{pages}{632--672}
  (\bibinfo{year}{2012}).

\bibitem{ying2021transformers}
\bibinfo{author}{Ying, C.} \emph{et~al.}
\newblock \bibinfo{title}{Do transformers really perform badly for graph
  representation?}
\newblock \emph{\bibinfo{journal}{Advances in Neural Information Processing
  Systems}} \textbf{\bibinfo{volume}{34}} (\bibinfo{year}{2021}).

\bibitem{vaswani2017attention}
\bibinfo{author}{Vaswani, A.} \emph{et~al.}
\newblock \bibinfo{title}{Attention is all you need}.
\newblock \emph{\bibinfo{journal}{Advances in neural information processing
  systems}} \textbf{\bibinfo{volume}{30}} (\bibinfo{year}{2017}).

\end{thebibliography}




\end{document}